\documentclass{statsoc}
\RequirePackage{natbib}

\usepackage{xspace}
\usepackage{amsmath,  amssymb}
\usepackage{fancyhdr, fancybox, graphicx,psfrag,dcolumn,bm,accents, textcomp,natbib}
\usepackage{color}
\usepackage{stmaryrd}
\usepackage[mathlines]{lineno}
\usepackage[usenames,dvipsnames]{xcolor}

\newcommand{\betaJk}{\boldsymbol{\beta}_{\mbox{\tiny{$\mathbb{J}(k)$}}}}
\newcommand{\betanJk}{\boldsymbol{\beta}_{\mbox{\tiny{$/\mathbb{J}(k)$}}}}

\newcommand{\tY}{\tilde{\boldsymbol{Y}}}
\newcommand{\AJk}{\mathbf{A}_{\mbox{\tiny{$\mathbb{J}(k)$}}}}
\newcommand{\DJk}{\mathbf{D}_{\mbox{\tiny{$\mathbb{J}(k)$}}}}
\newcommand{\Dtauk}{\mathbf{D}_{\mbox{\tiny{$\tau^2_{k}$}}}}
\newcommand{\pia}{\pi_{\mbox{\tiny{$\mathcal{A}$}}}}

\usepackage{amsmath, amssymb}
\usepackage{graphicx,psfrag,dcolumn,bm,accents, textcomp}
\usepackage[colorlinks=true, urlcolor=blue, citecolor=blue, linkcolor=blue]{hyperref}
\usepackage{color}
\usepackage{algorithm}
\usepackage[noend]{algpseudocode}

\usepackage{lineno}

\newcommand{\Jk}{\mathbb{J}(k)}
\newcommand{\betanj}{\boldsymbol{\beta}_{/j}}

\newcommand{\XinJk}{X_{\mbox{\tiny{$i,/\mathbb{J}(k)$}}}}

\newcommand{\tXJk}{\tilde{\boldsymbol{X}}_{\mbox{\tiny{$\mathbb{J}(k)$}}}}
\newcommand{\tXnJk}{\tilde{\boldsymbol{X}}_{\mbox{\tiny{$/\mathbb{J}(k)$}}}}
\newcommand{\lambdaa}{\lambda_{\mbox{\tiny{$\mathcal{A}$}}}}
\newcommand{\lambdad}{\lambda_{\mbox{\tiny{$\mathcal{D}$}}}}

\newcommand{\llasso}{\emph{lasso}\xspace}

\title[Group Bayes Selection]{Algorithmic Bayesian Group Gibbs Selection}

\author[Alan Lenarcic {\it and William Valdar}]{Alan Lenarcic}
\address{Department of Genetics, University of North Carolina at Chapel Hill, USA.}
\email{lenarcic@post.harvard.edu}

\author{William Valdar}\email{william.valdar@unc.edu}
\address{Department of Genetics, University of North Carolina at Chapel Hill, USA.}

\begin{document}
\bibliographystyle{chicago}

%%\linenumbers
\begin{abstract}
Bayesian model selection, with precedents in \cite{GeorgeMcCulloch} and \cite{AbramovichFirstPosteriorMedian:MR1649547}, support credibility measures that relate model uncertainty, but computation can be costly when sparse priors are approximate.  We design an exact selection engine suitable for Gauss noise, t-distributed noise, and logistic learning, benefiting from data-structures derived from coordinate descent lasso.  Gibbs sampler chains are stored in a compressed binary format compatible with Equi-Energy \citep{KourZhoWong} tempering.   We achieve a grouped-effects selection model, similar to the setting for group lasso, to determine co-entry of coefficients into the model.  We derive a functional integrand for group inclusion, and introduce a MCMC switching step to avoid numerical integration.  Theorems show this step has exponential convergence to target distribution.  We demonstrate a role for group selection to inform on genetic decomposition in a diallel experiment, and identify potential  quantitative trait loci in $p>40K$ Heterogenous Stock haplotype/phenotype studies. \end{abstract}

\section{Introduction}\label{Sec:Introduction}
   Linear model selection is used to reduce large multivariable regressions when there is little guidance over which explanatory variables are important, but that many are likely of negligible effect.  Competing methods are many and varied in estimator and algorithmic structure.  L1 penalized techniques related to the \llasso~\citep{Tibsh:Lasso} have appealing theoretical performance, and algorithms designed for \llasso  serve as building blocks for other penalty formulations \citep{Wang01082007,Zou:ElasticNet,Hui:AdaptLasso,zhang2010,candes2007}.  The canonical example for grouped, random effects is the group-\emph{lasso}~\citep{GroupLassoArticle}.

   Bayesian methods can incur heavier computational costs.  Often the prior is only semi-sparse, such as in such as~\cite{IshSpikeAndSlab} where coefficients are suppressed to a region near-to, but not exclusively, zero.  MCMC techniques resulting in truly sparse selection  have been to referred to as type ``Bayes-B'' or ``Bayes-C'' in the field of population genetics~\citep{MeuwissenBayesB}.  As shown with the fixed-effects sparse single-marker BSLMM model in ~\cite{StephensHSZhouEtAl} or in a non-sparse $p \approx 200$ group model~\cite{MALLICK2017115}, Bayesian technology is being driven to conduct Genome Wide Association Studies (GWAS) such as those on haplotype genomes where regions of the genome are descended from three or more original strains.  These studies drive a need for posterior selection measures choosing between multiple grouped random-effects.
	
	 Bayes techniques produce credibility intervals that give information about the precision of a measurement and instruct the user to either investigate features of high credibility, or to collect additional data to improve understanding in regions of uncertain posterior.  Credibility intervals, especially for the purpose of model selection, do not have objective or universal frequency coverage of a true value under all cases.  For instance, if the true value of $\beta_{j}/\sigma$ is $1 \times 10^{-9}$, a credibility model might classify $\beta_{j}$ as zero, since its effective contribution is so small as to be unobservable.  If the Bayesian theory reaches a posterior value $\mathcal{P}(\beta_{j} = 0) > .999$, this conclusion may be technically wrong, but useful in practice, as it supports an informed decision to ignore a negligible parameter.
   
   Here we implement a sparse Gibbs sampler~\citep{Gelfand1992}, first used in~\cite{Lenarcic413}, designed foremost for mixed effects and group selection, where groups of random effects  should be selected together.  We begin by detailing our implementation for fixed-effects, using analytic collapsed samples and data-structures suggested from Coordinate Descent~\citep{Tib:GLMNET} \llasso.  We store our sampler draws in a compressed binary format to ease the difficulties of recording large-$p$ Gibbs sampler chains.  This format is strongly compatible with an Equi-Energy \citep{KourZhoWong} tempering that allows MCMC to escape modes separated by regions of low probability.  We dynamically populate and reweight segments of the design-matrix to sample for $t$-noise and robit regression.
   
   To achieve group selection we generalize our fixed-effects scheme into a comparison between two integrable densities.  This density can be represented as a long-tailed, single-dimensional function potentially having multiple modes depending upon the eigenstructure of the subgroup. 
   Numerical integration of this function is difficult, so we introduce a new MCMC switching procedure based upon a comparison between bounding densities.  Since the switching-sampler is non-adaptive, theoretical mixing times of the sampler can be established.
   
   Simulation on fixed and grouped effects models shows competitive point estimation against common techniques, even when selection prior information is weak or wrong.  Augmenting the sampler  with non-sparse draws for credibility intervals helps intervals to cover parameters with realistic frequentist coverage, even for near zero features.  Our estimation method for the first-generation cross diallel experiment allows decomposition of the response into classes relating to modes of genetic inheritance.  We also show results on a publicly available Heteregenous Stock rat dataset, exploring models that suggest multiple quantitative trait loci (QTLs).

\section{Sparse Gibbs Sampling for Un-grouped Variables} \label{Sec:SparseGibbsUnGrouped}
 We begin with a common, intuitive, independent prior for coefficients $\beta_{j}$:
\begin{equation} \begin{split} B_{j} & \sim \mbox{Bernoulli}( \pia )\\ \beta_{j} & \sim N(0, B_{j} \times \tau^2_{\mbox{\tiny{F}}} )\mbox{, } \end{split}\label{SelectionFixed} \end{equation}
where a latent indicator $B_{j}$, $1$ or $0$ determines the active state of $\beta_{j}$.  ``$\pia$'' the prior probability of activation could be a global parameter (where $\pia \sim \mbox{Beta}(a,b)$ is a conjugate hyper-prior) or assigned with different strengths to individual $\pia(j)$ based upon experimental assumptions.  The variance of the ``On''-density $\tau^2_{\mbox{\tiny{F}}}$, reflecting the dispersion of fixed effects, could be set to a global value or  also weighted specific to coordinate, $\tau^2_{\mbox{\tiny{F}}}(j)$, which can allow for longer-tailed active priors.  If $\tau^2_{\mbox{\tiny{F}}}(j) \sim   \sqrt{ \nu / \chi^2_{\nu}} \tau^2_{0}$, then conditional on $\tau^2_{0}$, $\beta_{j}$ would have a long-tailed $t$ distribution prior, where $\tau^2_{0} \sim \mu_{\tau}\alpha_{\tau}/\mbox{Gamma}[\alpha_{\tau}]$ would be a possible hyperprior for the global dispersion parameter.
  This approach was conceptually first introduced by ~\cite{GeorgeMcCulloch}, but the first truly sparse implementations were designated a ``Bayes-B'' method in \cite{MeuwissenBayesB}, which adjusted model size through Metropolis-Hastings.

A typical approximation to this distribution is to consider two dispersions $\tau^2_{\mbox{\tiny{$\mathcal{A}$}}} >> 1$ and $\tau^2_{\mbox{\tiny{$\mathcal{D}$}}} << 1$ for the active and deactive prior cases of $\beta_{j}$, as used in \cite{GeorgeMcCulloch} and ~\cite{IshSpikeAndSlab}.  In this case, the posterior is a mixture of two distributions with a tall ``Spike'' at zero representing the active probability.  The benefit of this prior is that a Gibbs sampler may draw $B_{j} | \beta_{j}, \tau^2_{F}(j)$ conditional on current  estimate $\beta_{j}$.  The difficulties of this prior stem from a lack of true sparsity.  Since the discontinuous CDF is now approximated by a smooth derivative near zero, one must gauge carefully the adequate smoothing size of $\tau^2_{\mbox{\tiny{$\mathcal{D}$}}}(j)$, based upon sample size $n$, $p$, and other information, . 

Let us return to Equation ~\ref{SelectionFixed} and set zero-width $\tau^2_{D}=0$.   On first impression,  sampling $B_{j}| \beta_{j} = \beta_{j}^{(t)}$ seems impossible.  If last draw $\beta_{j}^{(t)} = 0$ then $B_{j}^{(t)} = 0$, but if $\beta_{j}^{(t)} \neq 0$ then $B_{j}^{(t)} = 1$ must be sampled.  Instead, draw $B_{j}^{(t)}$ from a  collapsed sample.  The collapsed posterior, $B_{j} | \vec{Y}, \boldsymbol{\beta}_{/j}$, is the distribution of $B_{j}$ using coefficient values for all $\beta_{j'}$ except for $\beta_{j}$.  Let $\mathcal{P}(\cdot|\cdot)$ represent the posterior, and $\mathcal{P}'(\cdot|\cdot)$ be an un-normalized proportion to the posterior. To integrate $\beta_{j}$ out of the posterior:
\begin{equation} 
  \begin{split} 
	\int_{-\infty}^{\infty} 
	   \mathcal{P}'( & B_{j},  \beta_{j}  |  \boldsymbol{\beta}_{/j},  \sigma^2, \tau^2_{F}(j), Y\ldots) d\beta_{j} \\ &
	\propto 
	   \int_{-\infty}^{\infty} e^{-\sum_{i}(Y_{i} - \mathbf{X}_{i,/j} \boldsymbol{\beta}_{/j} 
		                                            - X_{ij}\beta_{j})^2 / (2\sigma^2)} 
			      e^{-B_{j} \beta_{j}^2/(2\tau^2_{F}(j))} \left( 2 \pi \tau^2_{F}(j)\right)^{-.5 B_{j}}d \beta_{j} \\
& \propto  \frac{e^{-\frac{(\sum_{i}Y_{i} - \mathbf{X}_{i/j}\boldsymbol{\beta}_{/j})^2}{2\sigma^2}}}{\sqrt{ 2\pi \tau^2_{F}}^{B_{j}} } 
     \int_{-\infty}^{\infty} e^{\frac{\beta_{j}}{\sigma^2}\sum_{i} X_{ij}
		       (Y_{i} - \mathbf{X}_{i,/j} \boldsymbol{\beta}_{/j})} 
					 e^{-.5B_{j}\beta_{j}^2
					 \left( \frac{\sum_{i}X_{ij}^2}{\sigma^2} +\frac{1}{\tau^2_{F}(j)} \right)} d\beta_{j}  \\
& \propto \left( \frac{\tau^2_{F}(j)\sum_{i}X_{ij}^2}{\sigma^2 } + 1\right)^{-.5B_{j}}\exp\left\{.5 B_{j} \frac{ \left(\mathbf{X}^{T}_{,j}(Y - \mathbf{X}_{/j} \betanj)\right)^2}
  {\sigma^2 \sum_{i} X_{ij}^2 + \frac{ \sigma^4}{\tau^2_{F}(j)} } \right\} \mbox{.}
	\end{split} 
\label{ConditionalDistribution} 
\end{equation}
At the end of~\ref{ConditionalDistribution} only the terms proportional to $B_{j}$ are retained.  A future draw of:
\begin{equation}  B_{j} \sim \mbox{Bernoulli} \left( \frac{ \pia(j) \mathcal{P}'(B_{j}=1| \boldsymbol{\beta}_{/j}) }
{\pia(j) \mathcal{P}'(B_{j}=1| \boldsymbol{\beta}_{/j}) + (1-\pia(j)) \mathcal{P}'(B_{j}=0| \boldsymbol{\beta}_{/j})} \right) \label{BernoulliBj2} \end{equation}
can then be made, iterating through coordinates $j$.  

Equation~\ref{ConditionalDistribution} suggests two statistical quantities that determine  $B_{j}$ activation: the sum of squares $\sum_{i} X_{ij}^2$ and the correlated residual $\sum_{i} X_{ij}(Y_{i} - \mathbf{X}_{i,/j} \boldsymbol{\beta}_{/j})$.  As described in ~\cite{Friedman:CoordDesc} these quantities also appear in the Coordinate Descent \llasso (CDL) algorithm, and a Bayesian Gibbs sampler can gain efficiency by adopting the same allocation scheme.  Similar to CDL, to efficiently update a vector $\mbox{XTResid}$ in $\mathbb{R}^{p}$, defined:
\begin{equation} \mbox{XTResid} = \mathbf{X}^{T} \left( \vec{Y} - \mathbf{X} \boldsymbol{\beta} \right), \label{XtResid} \end{equation}
it is required to calculate the $p\times p$ ``sum of squares'' matrix $\mathbf{X}^{T} \mathbf{X}$ for all columns $j$ where $\beta_{j}$ is non-zero.  If $\boldsymbol{\beta}$ is sparse, then this is only a subset of all of $\mathbf{X}^{T} \mathbf{X}$.  So, as the sampler begins, only a few columns of $\left\{\mathbf{X}^{T}\mathbf{X} \right\}_{,j}$, for $j$ in the active coefficients, need to be stored in memory.   When other coordinates join the active set, additional columns are added and  memory is occasionally allocated in blocks to handle future columns.  If some coordinates $j$ are present in early models but do not return after many chain iterations, their memory columns are freed.  After every mutation of $\boldsymbol{\beta}^{(t)}$ to $\boldsymbol{\beta}^{(t+1)}$ is made, the vector $\mbox{XTResid}$ can be updated in $\mathcal{O}(p \times \| \mathcal{A}^{(t+1)} \|)$ time, where $\| \mathcal{A}^{(t+1)}\|$ is the size of the nonzero set of $\boldsymbol{\beta}^{(t+1)}$.  Prudent adjustment of $\mbox{XTResid}$ is made after the update of each coordinate $\beta_{j}^{(t)}$ to $\beta_{j}^{(t+1)}$.  With $\mbox{XTResid}_{j+1}$ updated,  Equations~\ref{ConditionalDistribution} and~\ref{BernoulliBj2} can decide the next $B_{j+1}^{(t+1)}$, $\beta_{j+1}^{(t+1)}$, and the loop continues through all coordinates.  Such a parsimonious, sparse sampler could potentially tackle large $p \geq 100,000$ on a small machine using a single CPU core, if issues of storage, mixing, and non-Gaussian noise are also  tackled.

 For extended models: logistic, robit, and $t$-regression, these require a weighted sum $\widetilde{ \mathbf{X}^{T} \mathbf{X}}$ defined as having elements $ \{\widetilde{ \mathbf{X}^{T} \mathbf{X}} \}_{jk} = \sum_{i} X_{ij} X_{ik} w_{i}^{(t)}$ for a vector of weights $\vec{w}^{(t)} \in \mathbb{R}^{p}$ giving unequal weight to each datapoint and  reweighted during each iteration.   $\widetilde{ \mathbf{X}^{T} \mathbf{X}}_{,j}$ need only be updated on current active columns.  Weights of this form are relevant for non-Gaussian noise. For $t_{\nu}$-distributed long-tailed noise $Y_{i}-\mathbf{X}_{i,}\boldsymbol{\beta} \equiv \varepsilon_{i} \sim \frac{\sigma}{\sqrt{w_{i}}} N(0,1)$ where $w_{i} \sim \mbox{Gamma}(\nu/2) / (\nu/2)$.  For a binomial-family generalized linear model, consider observed $Z_{i} \sim \mbox{Bernoulli}( p(X_{i}) )$ for $p(X_{i}) = \Phi^{-1}_{\nu}( \mu + \mathbf{X}_{i,} \boldsymbol{\beta})$.  This $\Phi^{-1}_{\nu}(x)$ is an inverse CDF from a desired distribution: the Gaussian inverse CDF, $\Phi^{-1}(x)$, in the probit case, or the inverse CDF of a $t$-distribution d.f. $\nu$, $\Phi^{-1}_{\nu}(x)$, in robit regression.   To update the weight $w_{i}^{(t)}$ for $t_{\nu}$-noise, sample:
\begin{equation} w_{i}^{(t)} | \boldsymbol{\beta}^{(t)}, \sigma^2 \sim   \frac{2\sigma^2\mbox{Gamma}\left((\nu+1)/2\right)}{\sigma^2\nu +(Y-\mathbf{X}_{i,} \boldsymbol{\beta}^{(t)})^2} \mbox{.} \label{wVectorDfT} \end{equation}
For robit regression,  draw a latent $Y_{i}$ given $\boldsymbol{\beta}$:
\begin{equation}  Y_{i} | \boldsymbol{\beta} \mbox{ \& } (Z_{i}=1) \sim \mathbf{X} \boldsymbol{\beta} +\sigma\mbox{Truncated-T}\left(\nu, - \mathbf{X} \boldsymbol{\beta}/\sigma, \infty\right) \label{weightedTDrawVector} \end{equation}
and the opposite tail, ($-\infty$ to$- \mathbf{X} \boldsymbol{\beta}/\sigma$)  if $Z_{i}=0$.  Slice-sampling is used to achieve the $t$-tail distribution draw in Equation~\ref{weightedTDrawVector}.  Once $Y_{i}^{(t)}$ is drawn in the robit model, the robit update for $w_{i}^{(t)}$ can come from the same Equation~\ref{wVectorDfT}. For logistic regression, following \cite{MudholkarAndGeorgSeae}, we approximate with a robit of degrees of freedom 9 and $\sigma^2= \frac{7 \pi^2}{27}$ to set variance and kurtosis equivalent to the logit, and use importance weights to correct the mean and confidence intervals.  Rebalancing $\tilde{Y}^{(t)}$ and $\tilde{X}^{(t)}$ can significantly slow runtime; maintaining a sparse $\hat{\boldsymbol{\beta}}$ through a heavy penalty, and approximating with $w_{i}^{(t)}$ remaining constant for several draws, can relieve some of the impact.  
Let the set $\mathcal{A}^{(t)}$ be the coordinates of $\boldsymbol{\beta}$ such that $B_{j}^{(t)} = 1$.  Once $\mathcal{A}^{(t)}$ has been decided for an iteration, assuming that $\mathcal{A}$ is much smaller than $1, \ldots, p$, a draw: \begin{equation} \boldsymbol{\beta}_{\mathcal{A}^{(t)}}^{(t)} \sim N\left( (  \mathbf{Q}_{\mathcal{A}})^{-1} \widetilde{\mathbf{X}_{\mathcal{A}} Y}, {\sigma^2}^{(t)} \left[ \mathbf{Q}_{\mathcal{A}}\right]^{-1} \right) \label{BetaDraw} \end{equation}
is taken through Cholesky decomposition.   $\mathbf{Q}_{\mathcal{A}} \equiv \widetilde{\mathbf{X}_{\mathcal{A}^{(t)}}^{T} \mathbf{X}_{\mathcal{A}^{(t)}}} + \mathbf{D}_{\tau^2_{F}}$ is the appropriately weighted square of active covariates plus a diagonal matrix proportional to $\sigma^2/\tau^2_{F}(j)$.  If $\|\mathcal{A}\| > 400$, as the $\mathcal{O}(\|\mathcal{A}\|^{3})$ Cholesky decomposition becomes difficult, then invert $\mathbf{Q}_{\mathcal{A}}$ in blocks.  The  Gibbs sampler would update a subset of active coefficients $\mathcal{A}_{1}$, and then update additional subsets until complete.
Again, having $\mbox{XTResid}$  allows the calculation of 
   $X_{\mbox{\footnotesize{$\mathcal{A}_{1}$}} }
	      \left(Y - \mathbf{X}_{,\mbox{\footnotesize{$\mathcal{A}_{2}$}}}
				\boldsymbol{\beta}_{\mbox{\footnotesize{$\mathcal{A}_{2}$}}}\right)$, which
appears in Equation~\ref{BetaDraw} for  the conditional mean.

\subsection[Efficient Gibbs Chain Storage]{Efficient Storage}\label{Sec:EfficientStorage}
  Gibbs samplers can require significantly more memory than maximization methods.  
	While costs of storage of $S$ chains worth of $T$ samples of a $p$ length 
	  $\boldsymbol{\beta}$-vector are manageable for medium sized $p$, 
		allocation for chains becomes impracticable for cases  $p > 10,000$.  
	Chain-thinning may lead to more efficient samples, but since our sampler
	  draws from the multivariate $\boldsymbol{\beta}| \sigma^2, Y$ posterior, 
		draws  already show low autocorrelation.  
	So while every sample is useful, it is preferable to free those samples 
	  from RAM.  	Allocating memory for draws of all coefficients $\beta_{j}$ is inefficient,
	  given the presumed sparsity of the draws.  
	For large $p$, we expect a significant majority of $\hat{\beta}^{(t)}_{j} = 0$ 
	  on a given draw, but active set membership  will be changeable and somewhat unpredictable.  Before analysis, the subset of interesting coefficients is unknown, and is not of fixed size.  But, in selection settings, we do hope that at a draw, $(t)$, the active set size $\| \mathcal{A}^{(t)} \|$ is much less than $p$.  
  
  We implement a binary-file buffer storage scheme  to reduce RAM-costs and simplify appending to files.  Two buffers are stored: one a length $\mathcal{N}$ linear array of double-point memory, ``$D[]$'', storing a sequence of the non-zero coefficients of $\beta$, and the second a matching length $\mathcal{N}$ 
	linear array of integer-valued index locations, ``$I[]$''.  As a sample $(t)$ is taken, with most $\hat{\beta}^{(t)}_{j}=0$, the non-zero $\hat{\beta}^{(t)}_{j}$ are stored in the buffer in increasing order of $j$.  Leading off each write of state (t) to the buffers,  the iteration identifier $-t$ is stored in $D[]$, and at the corresponding position, in buffer $I[]$ we store 
		$-\| \mathcal{A}^{(t)} \|$ or length of the active set (convention of a negative sign to signal this is not a coefficient).  When these buffers are filled near $\mathcal{N}$,  they are appended to files in disk storage and the buffers are wiped.  
	
	For later selection structures, where $\beta_{j}$ are arranged in groups set non-zero by different $\tau^2_{k}$ parameters, a sparse comparable buffer system for storing the $\boldsymbol{\tau}^2$ vector is implemented.  Buffers for the Rao-Blackwellized $\mathcal{P}( B_{j}=1| \boldsymbol{\beta}_{/j}^{(t)}) \approx \mathbb{E} \left[ B_{j}^{(t+1)} | \boldsymbol{\beta}^{(t)} \right]$ are stored in a separate file to calculate posterior Marginal Inclusion Probabilities (MIPs)  for each coefficient $\beta_{j}$.  Furthermore, after each draw a buffer $\mathcal{P}'( \boldsymbol{\Theta}^{(t)} |Y)$ stores posterior probabilities (up to unknown integration constant), of each draw of of coefficients $\boldsymbol{\Theta}^{(t)} = \left\{ \boldsymbol{\beta}^{(t)}, \boldsymbol{\tau}^{(t)}, \sigma^{(t)}, \boldsymbol{\pi}_{\mbox{\tiny{$\mathcal{A}$}}}^{(t)}, \mathbf{w}^{(t)} \right\}$.  
  
  These chains saved  for later analysis, we can recover R-Package Coda~\citep{PlummerCODA} 
	  ``{\small{\verb.MCMC.}\normalsize}'' objects including just a subset of coefficients $\boldsymbol{\beta}_{j}$.  We retrieve $\beta_{j}$ chains only with largest MIP: $\frac{1}{T}\sum_{(t)} B_{j}^{(t)} \approx \frac{1}{T} \sum_{t} \mathcal{P}( B_{j}=1| \boldsymbol{\beta}_{/j}^{(t)})$.  This implementation detail may seem irrelevant to the statistical properties of the sampler, but our storage choice serves a role in later discussions on Equi-Energy tempered sampling. 		Further, space-requirements for Gibbs samples can be a drawback to wider use of Bayes methods in the large $p$ realm, but sparse-packing can address this issue.
\subsection{Equi-Energy Sampling}
  When multiple disjoint models each adequately predict $Y$, a sampler may become stuck in a local mode.  As illustration, consider a low-noise case, where a strong predictor covariate $X_{,j_{1}}$ is highly correlated, and thus replaceable, with another vector $X_{,j_{2}}$.  Drawing more than a single coordinate of $\boldsymbol{B}^{(t)}$ at a time may alleviate this problem, exploring distant models can still be difficult.  To search a larger model-space, additional chains from higher temperatures $\mathcal{P}'(\boldsymbol{\Theta})^{1/T}$ can be taken and merged into the chains from $T=1$ base temperature.  
  
  Rather than run a system of parallel tempering, as per~\cite{GeyerTempering}, we find more compatibility with the~\cite{KourZhoWong} ``Equi-Energy'' (EE) tempering.  Here, first the highest temperature $T_{1}$ chains of short length are run to explore the range of the posterior.  These chains are sorted and stored in order of $\mathcal{P}'(\boldsymbol{\Theta}|Y)$ posterior density (up to unknown constant) values.  Then chains at the second highest temperature $T_{2}$ are begun.  After a period of every $\Delta_{t} = 50$ or so iterations, a merge step is proposed from previous draws from $T_{1}$.  To draw a merge,  consider parameter set $\boldsymbol{\Theta}^{(t)}$ at the current iteration and estimated value $\mathcal{P}'(\boldsymbol{\Theta}^{(t)}|Y)$.  Then seek $\boldsymbol{\Theta}^{\mbox{\tiny{prop}}}$  from the draws from $T_{1}$  within $\left|\mathcal{P}'(\boldsymbol{\Theta}^{(t)}|Y) - \mathcal{P}'(\boldsymbol{\Theta}^{\mbox{\tiny{prop}}}|Y) \right| < \varepsilon$.  
	If, for small $\varepsilon$ many $\boldsymbol{\Theta}^{\mbox{\tiny{prop}}}$ values can be found, a uniform draw from this set is selected to replace $\boldsymbol{\Theta}^{(t)}$ in the current chain at $T_{2}$.  For cases where $\varepsilon$ need be large, a small Metropolis-Hastings correction:
  \begin{equation} \mbox{Merge if: \hspace{2mm}}  \mathcal{P}'\left( \boldsymbol{\Theta}^{\mbox{\tiny{prop}}} |Y \right) / \mathcal{P}'\left( \boldsymbol{\Theta}^{(t)} |Y \right) \geq \mbox{Unif}(0,1)  \label{MergeEquation} \end{equation}
  determines the merge. 	The EE sampler can quickly and frequently leave the current mode.  Abandoning a previous mode places extra pressure on Coordinate-Descent dynamic memory, but, for lower temperatures,  the number of actively needed columns of $\mathbf{X}^{T} \mathbf{X}$  should be reasonable.   
		In EE sampling, chains in the highest temperature can be drawn well in advance, on different days from different machines.  Though two temperatures cannot be drawn in parallel, several chains at the same temperature can be drawn independently.  This EE-sampler should benefit from the sparse-storage method explained in Section~\ref{Sec:EfficientStorage}. Files at higher temperatures can be sorted by likelihood, enabling efficient random access.  
 \section{Group Sampling}\label{Sec:GroupedVariables}
 Building  on Section~\ref{Sec:SparseGibbsUnGrouped}'s techniques  for  fixed-effects, we construct a method for random-effects. Consider a prior for groups of coordinates $j \in \Jk$:
   \begin{equation} \begin{split} \tau^2_{k} & \sim \mbox{Bernoulli}\left( \pia(k)\right) \times \frac{\tau^2_{0} \nu_{\tau}}{\chi^2_{\nu_{\tau}}} \\
      \beta_{j} & \sim N(0, \tau^2_{k} ) \mbox{, for all $j \in \Jk$ \hspace{3mm} }. \end{split} \label{GroupPrior} \end{equation}
 In Equation~\ref{GroupPrior},  $\tau^2_{k}=0$  with prior probability $\pia(k)$ and, if not, has an inverse chi-squared prior.  This prior can be used with constrained coefficients, as we show in Section~\ref{SectionIdentifiabilityConstraint}.  Let the length of each group, $J_{k} \equiv\|\Jk\|$.

Generalizing the collapsed sampler case described in Section~\ref{Sec:SparseGibbsUnGrouped}, we seek to perform a draw of $\tau^2_{k} | \vec{Y}, \betanJk$ where the values of $\betaJk$ for coordinates in $\Jk$ have been integrated out of the posterior.  Unlike Equation~\ref{ConditionalDistribution}, this requires a multivariate integration in space $\mathbb{R}^{J_{k}}$.  We can make this integration tractable with two steps.  First, in Section~\ref{DeriveActiveDistribution} we rotate $\mbox{XTResid}$ by eigenvalues of $\{ \mathbf{X}^{T} \mathbf{X} \}_{\Jk,\Jk}$ to reduce the problem to a $\mathbb{R}^{1}$ function, $f_{2;k}(\tau^2_{k} )$, representing an unintegrated density.  Integrating $f_{2;k}(\tau^2_{k} )$ with accuracy is still a slow process.  So secondly, to speed up this step some $100$ times, we propose a bounded-density Markov chain in Section~\ref{RandProbMarkovChain}, for which we prove theoretical convergence in Section~\ref{Sec:JRSSTheoreticalResults}.
\subsection{Recentering a group prior for an identifiability constraint}\label{SectionIdentifiabilityConstraint}
While useful for many cases, the prior in Equation~\ref{GroupPrior}  may inadequately encode known restrictions.  In genetics, there may be 8 possible haplotypes ``A'', ``B'', \ldots ``H'' at a given loci and a regression may be of the form:
\begin{equation} Y_{i} = \mu + 1_{X_{i} = A} \beta_{A} + 1_{X_{i}=B} \beta_{B} + \ldots 1_{X_{i}=H} \beta_{H} + \varepsilon_{i} \mbox{. }\label{HaplotypeRegresion} \end{equation}
It would be of interest to the researcher to know whether $|\beta_{A}| + |\beta_{B}| + \ldots |\beta_{H}|$ is non-zero.  The design matrix, $\mathbf{X}$,  will have linearly dependent columns, since the presence of the haplotype ``H'' is determined by the absence of the first seven.   In general, a genetics design matrix can have a constraint $\sum_{j\in \Jk}\mathbf{X}_{ij} = \zeta$ for all $i$ for columns group $\Jk$.
When an independence prior of the form $\beta_{k} \sim N(0,\tau^2_{k})$ is applied to this haplotype region,  the posterior becomes proper and a posterior mean exists.  However, credibility intervals for $\beta_{A}$ through $\beta_{H}$ will be wide due to a non-identifiability against $\mu$.    As a solution consider a constraint on the prior:
\begin{equation}\begin{split} \beta_{A} + \beta_{B} + \ldots & + \beta_{H}  = 0 \\
    \beta_{A}, \beta_{B}, \ldots, \beta_{H} \sim & \mbox{ marginally } N(0, \tau^2_{k}) \mbox{.} \end{split} \label{constrainedPrior} \end{equation}
The overall mean, $\mu$, in regression equation~\ref{HaplotypeRegresion} is no longer confounded with the sum of the haplotype regressors.  When group regression for group $k$ has $J_{k}$ linearly dependent columns of this form we construct a transformation matrix:
\begin{equation}  \mathbf{M} = \left[ \begin{array}{ccccc} c & -b & -b & \ldots & -b \\
         -b & c & -b & \ldots & -b \\
				 -b & -b & c & \ldots & -b \\
				 \ldots & \ldots & \ldots & \ldots & \ldots \\
				  \ldots & \ldots & -b & -b & c \\
					-d & -d & -d & \ldots & -d \end{array} \right] \mbox{.} \label{MTransformation} \end{equation} 
Where $\mathbf{M}$ is a matrix with $J_{k}$ rows, but $J_{k}-1$ columns.  
If the $d = 1/\sqrt{J_{k}-1}$, $b = \frac{-1 + \sqrt{J_{k}}}{(J_{k}-1)^{3/2}}$ and 
  $c = (J_{k}-2)b + d$, then the matrix $\mathbf{M}$ has columns that naturally sum to 
	zero, but also where $\sum_{\nu} M_{\eta\nu}^2 = 1$ for all rows $\eta$.   $\mathbf{M}\mathbf{M}^{T} \vec{v} = \frac{J_{k}}{J_{k}-1} \vec{v}$ for $J_{k}$ length vectors $\vec{v}$ that sum to zero.
	
If $\beta_{k}'$ is a $J_{k}-1$ length iid $N(0,\tau^2_{k})$ a-prior vector, then $\beta_{k} \equiv \mathbf{M} \beta_{k}'$ will sum to zero, but marginally  each element is $N(0,\tau^2_{k})$.

Mixed effects methods which try to minimize the sum of a likelihood and a penalty, such as in group-\emph{lasso}  or non-sparse REML~\citep{Bartlett:Reml}, generally suggest $\sum_{k} \beta_{k}=0$ constraint.  But a Bayesian sampler must enforce this constraint through some prior,  and with our choice being Equation~\ref{MTransformation}. 

The group procedures following in this section~\ref{Sec:GroupedVariables} can be assumed to be performed unconfounded on transformed  design columns $\mathbf{X}_{k'}' = \mathbf{X}_{k} \mathbf{M}$ and posterior inclusion of transformed groups slightly downgraded by the change of measure.

 \subsection{Deriving Active Distribution for $\tau^2_{k}$}\label{DeriveActiveDistribution}
   Begin combining prior and likelihood
for a form proportional to the posterior:
 \begin{equation} \begin{split} 
  \mathcal{P}  (& \boldsymbol{\beta},   \boldsymbol{\tau}^2 | Y, X, \sigma^2 ) \propto \mathcal{P}' ( \boldsymbol{\beta},   \boldsymbol{\tau}^2 | Y, X, \sigma^2 ) \equiv
  \\ \equiv &
   \frac{1}{\sqrt{2\pi \sigma}^{n}} 
	  \exp \left\{ -\frac{1}{2\sigma^2} 
		       \left( \mathbf{Y} - \mathbf{X} \boldsymbol{\beta} \right)^{T}  
           \left( \mathbf{Y} - \mathbf{X} \boldsymbol{\beta} \right) -     
     \sum_{k} \sum_{j \in \mathbb{J}(k)} \frac{\beta^2_{j}}{2\tau^2_{k}} \right\} 
     \prod_{k=1}^{K} \frac{p(\tau^2_{k})}{\sqrt{2 \pi \tau^2_{k}}^{J_{k}}}\mbox{. } \end{split}
 \label{DensityOfY}
\end{equation}

 Now to integrate out $\betaJk$.  Define this function of interest, $f_{2;k}(\tau^2_{k})$, to be:
 \begin{equation}\begin{split} f_{2;k} & (\tau^2_{k})  \equiv \int_{\betaJk} \mathcal{P}'( \betaJk, \betanJk, \tau^2_{k}, \tau^2_{/k} | Y, X ) d \betaJk  \\
& = 
\int_{\boldsymbol{\beta}_{\mathbb{J}(k)}} e^{ -\left[\sum_{i=1}^{N} \frac{1}{2\sigma^2_{i}} ( Y_{i} - \boldsymbol{X}_{\mbox{\tiny{$i,/\mathbb{J}(k)$}}} \boldsymbol{\beta}_{\mbox{\tiny{$/\mathbb{J}(k)$}}} -
    \boldsymbol{X}_{\mbox{\tiny{$i,\mathbb{J}(k)$}}} \boldsymbol{\beta}_{\mbox{\tiny{$\mathbb{J}(k)$}}} )^2-\sum_{j \in \mathbb{J}(k)} \frac{\beta_{j}^2}{2\tau^2_{k}} \right]} \frac{p(\tau^2_{k})}{\sqrt{ 2\pi \tau^2_{k}}^{J_{k}}}d \boldsymbol{\beta}_{\mathbb{J}(k)}. \end{split}
\label{f2taukDefined} \end{equation}   

Assume out-of-group parameters, $\betanJk$, are held constant at this step of the Gibbs sampler.   $p(\tau^2_{k})$ is a mixed prior $(1-\pia(k))\delta(\tau^2_{k}) + \pia(k) p_{2}(\tau^2_{k})$,  with finite measure for $\tau^2_{k}=0$.   Excluding zero, the integration  $\int_{0^{+}}^{\infty} f_{2;k}(\tau^2_{k}) d\tau^2_{k}$ is proportional to the posterior probability that $\tau^2_{k} > 0$.   ``$F_1$'', the integration at zero, $\int_{0}^{\infty} \delta(\tau^2_{k}) f_{2;k}(\tau^2_{k}) d \tau^2_{k}$ is the alternative.  We will reduce the analytical form for $f_{2;k}(\tau^2_{k})$ so that it can be computed in $\mathcal{O}(\mathbb{J}(k))$ time.

\
Define $\tilde{\boldsymbol{X}}$ as the weighted matrix such that 
  $\tilde{X}_{ij} \equiv \frac{1}{\sigma_{i}} X_{ij}$, and $\tilde{Y}_{i}$ such that 
	$Y_{i} \equiv \frac{Y_{i}}{\sigma_{i}}$.     
Let 
  $S^2 = \sum_{i} \left( \tilde{Y}_{i} - \tilde{\mathbf{X}}_{\mbox{\tiny{$i, /\mathbb{J}(k)$}}} \betanJk \right)^2 
	     = \sum_{i} \frac{1}{\sigma^2_{i}}\left( Y_{i} - \XinJk \betanJk \right)^2$, which is the same in $\tau^2_{k}$ on-state and off-state.  Also, let $\mathbf{D}_{\tau^2_{k}}$ be a diagonal matrix with diagonals proportional to $\frac{1}{\tau^2_{k}}$ and size of length $\betaJk$.  
 \begin{equation} 
   f_{2;k}(\tau^2_{k})   =\frac{e^{-S^2}p(\tau^2_{k})}{\sqrt{2\pi \tau^2}^{J_{k}}} \int_{\betaJk} e^{ 2 (\tY - \tXnJk)^{T}\tXJk\betaJk - \betaJk^{T} \left(\tXJk^{T} \tXJk + \mathbf{D}_{\tau^2_{k}} \right)\betaJk }d\betaJk \mbox{.} \label{FixedEquation}\end{equation}

Define $\tilde{\boldsymbol{R}} =  \tXJk^{T}\left(\tY-\tXnJk \betanJk \right)$.  Completion of the square reduces this to:
  \begin{equation}  f_{2;k}(\tau^2_{k})   =  \frac{e^{-S^2}p(\tau^2_{k})}{\sqrt{\tau^2}^{.5J_{k}}}\sqrt{ | \tXJk^{T} \tXJk + \mathbf{D}_{\tau^2_{k}} |^{-1} }e^{\frac{1}{2} \tilde{\boldsymbol{R}}^{T} \left(\tXJk^{T} \tXJk + \mathbf{D}_{\tau^2_{k}} \right)^{-1}\tilde{\boldsymbol{R}}} .\label{TauSqDens} \end{equation}
	
	A simpler form of this equation can be made by rotating through the eigen-decomposition: $\tXJk^T\tXJk = \AJk \DJk \AJk^{T}$ for right eigenvector matrix $\AJk$.  Since $\Dtauk = \mathbf{I} / \tau^2_{k}$ is diagonal with all terms equivalent to $\tau^2_{k}$, and hence $\Dtauk \AJk = \frac{1}{\tau^2_{k}} \AJk$, certain computations are possible:
  \[   \left( \tXJk^{T}\tXJk + \Dtauk \right)^{-1} = \AJk^{T}\left( \DJk + \Dtauk \right)^{-1}  \AJk \mbox{. }\] 
	 Thus the determinant is proportional to $\prod_{j=1}^{J(k)} D_{\mathbb{J}(k):j} + \frac{1}{\tau^2_{k}}$.  Further,:
  \[ \tilde{R}^{T} \left(\tXJk^{T} \tXJk + \mathbf{D}_{\tau^2_{k}} \right)^{-1}\tilde{R} = \mbox{tr} \left\{ \AJk \tilde{R} \tilde{R}^{T} \AJk^{T} \left( \DJk + \Dtauk\right)^{-1} \right\} \mbox{ .} \]
  	Define $\mathbf{R}_{A} \equiv \AJk \tilde{R} \tilde{R}^{T} \AJk^{T} $, then the sum is $\sum_{j \in \mathbb{J}(k)} \frac{\mathbf{R}_{A:j,j}}{ {\DJk}_{j,j} + \frac{1}{\tau^2_{k}}}$.  Computation of the diagonal $\mathbf{R}_{A:j,j}$ is quick when we realize each value is:
  \begin{equation}  \mbox{Diag}\left( \mathbf{R}_{A} \right)_jj =  \left[ \AJk^{T} \tXJk^{T}(\vec{Y} - \tXnJk \betanJk) \right]_{jj}^2  \mbox{ .}\label{DiagEquation} \end{equation}
 	Finally we express $f_{2;k}( \tau^2_{k}) $ as:
	\begin{equation} f_{2;k}(\tau^2_{k}) = 
	  p(\tau^2_{k})e^{-S^2} \times \prod_{j=1}^{J_{k}} 
		  \sqrt{\frac{1}{D_{\mathbb{J}(k):j} \tau^2_{k} + 1}}
		  \exp\left\{ +.5 \sum_{j=1}^{J_{k}}\frac{R_{A:j,j}}{D_{\mathbb{J}(k):j} + \frac{1}{\tau^2_{k}}}
	        \right\} \mbox{.} 
	\end{equation}
  Hence, eigensolvers can convert $\tau^2_{k}$'s marginal posterior to a $O(J_{k})$ function.  
	 Note that the factor in the exponential is positive, with limit $1$ as $\tau^2_{k} \rightarrow \infty$.  Excluding the prior, the likelihood portion of $f_{2;k}(\tau^2_{k})$ decays only at a ${\tau^2}^{-.5 J_{k}}$ rate.
  
\begin{figure}[htbp]
	\centering
		\includegraphics[width=1.00\textwidth]{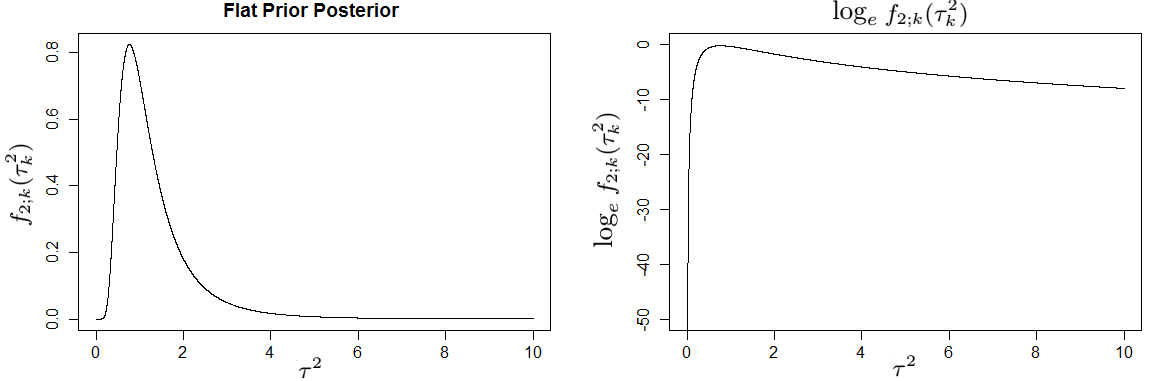}
	\caption{An example $f_{2;k}(\tau^2_{k}), \log_{e} f_{2;k}(\tau^2_{k})$ given $k=8$, flat-prior, and 
	  some representative eigenvalues.    
	This may have one or more modes, and a slow polynomial decay $\propto (\tau^2)^{-.5J_{k}}$.    
	In preventing large draws of $\tau^2$, even an inverse-chi-squared 
	  $\propto (\tau^2)^{-\nu/2-1} e^{-\bar{\tau}\nu/(2\tau^2)}$ would be a weak prior, 
		since this decays as a polynomial as $\tau^2$ gets large. 
	}
	\label{fig:BayesSpikeLikelihood}
\end{figure}

	Integration $\int_{0+} f_{2;k}(\tau^2_{k})d \tau^2_{k}$ calculates the conditional odds that coefficients in $\mathbb{J}(k)$ should be non-zero.  
		Without the benefit of possibly a GPU, or some accelerated quadrature, numerical integration can be slow, considering the  challenge of a long tail.  We propose a sampling-based integration in the next Section~\ref{RandProbMarkovChain}.

\subsection{Using Bounding Densities For a Convergent Sampler}\label{RandProbMarkovChain}

    We propose a novel method for calculating selection probabilities for the
				density in Equation~\ref{TauSqDens} and in further problems.  Without loss of generality we keep $F_{1}=1$,  and consider a function $f_{2}(x) \in \mathbb{R} \rightarrow (0+, \infty)$ defined to be $f_{2}(x) = F_{2} \times q_{2}(x)$ where $F_{2}$ is unknown, and $q_{2}(x)$ is the sampling density of $X$ (i.e. $\int q_{2}(x) dx = 1$).  Integration of $f_{2}(x)$, to solve for $F_{2}$, is still a slow process, though random samples $X_{2}$ from $q_{2}(x)$ are possible due to slice sampling.
				  We know that $\mathbb{E} \left[ \frac{q(x)}{f_{2}(X_{2})} \right] = \int \frac{q(x)}{f_{2}(x)} q_{2}(x) dx = \frac{1}{F_{2}}$, where $q(x)$ is any density with equivalent support to $q_{2}(x)$. 
We construct a Markov sequence, $W^{(t)}$, such that $\lim_{T\rightarrow \infty} \frac{1}{T} \sum_{t} W^{(t)}  = \frac{F_{2}}{1+F_{2}}$.  Let current state $W^{(t)}$ determine whether $W^{(t+1)}$ is drawn with probability $A^{(t)}$ if $W^{(t)}=1$, or $D^{(t)}$ if $W^{(t)}=0$.   $A^{(t)}, D^{(t)}$ are a sequence of draws which are independent of $W^{t}$ and all previous draws $A^{(t-i)}$, $D^{(t-i)}$. 

\begin{equation} \mathbb{P}(W^{(t+1)} | W^{(t)}, A^{(t)}, D^{(t)} ) = \begin{array}{c} \mbox{``De-Active''} \\ \mbox{``Active''} \end{array} : \left[ \begin{array}{cc} 1-D^{(t)} & D^{(t)}  \\
  1-A^{(t)} & A^{(t)}  \end{array} \right] \mbox{.} \label{SwitchingDistributon} \end{equation}

\begin{figure}[htbp]
	\centering
		\includegraphics[width=0.90\textwidth]{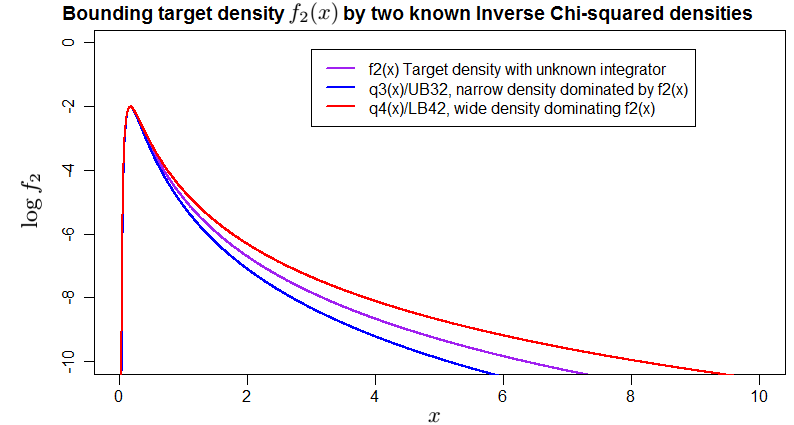}
	\caption{Consider $f_2(x)$ a density with unknown integrator $\int f_2(x) dx = F_{2}$.  However, we can readily find its maximum point, and we know its curvature and tail properties.  We can thus consider inverse-gamma densities that bound the function.}
	\label{fig:DensityBayesSpikeDemonstration1}
\end{figure}

   To specify $A^{(t)}, D^{(t)}$, we use upper and lower bounding functions  for $q_{2}(x)$ whose integrations are known, such as demonstrated in Figure~\ref{fig:DensityBayesSpikeDemonstration1}.  Bound $f_{2}(x)$ by  $f_{4}(x) = F_{4>2} q_{4}(x)$ and $f_{3}(x) = F_{3<2} q_{3}(x)$ with property $f_{3}(x) \leq f_{2}(x)$, and $f_{3}(x) \leq f_{4}(x)$ almost everywhere  (that $f_{4}(x)\succeq f_{2}(x)$ is optional, but $f_{4}(x)$ should dominate in some region).  $q_{3}, q_{4}$ are true density functions, such that $\int q_{3}(x) dx = \int q_{4}(x)dx = 1$.  The constant $F_{3<2}$ represents the highest lower bound for $f_{2}(x)/q_{3}(x)$, and $F_{4>2}$ works best if it is the lowest upper bound of $f_{2}(x) / q_{4}(x)$.
       
In our case of $f_{2;k}(x=\tau^2_{k})$ , inverse-gamma densities can be chosen for $q_{3}(x), q_{4}(x)$.  Frequently $f_{2}(x)$ will be unimodal, so arg-max $x_{2}^{\mbox{\tiny{max}}}$ of $f_{2}(x)$ can be achieved through Newton methods or binary search.   Knowing the curvature $f_{2}''(x_{2}^{\mbox{\tiny{max}}})$ hints at what $q_{3}, q_{4}$ functions should be.  Also valuable is the tail constant: $d_{k}$ such that $ \lim_{x \rightarrow \infty} f_{2}(x)/x^{d_k} \rightarrow \mbox{Constant}$.  Choose a $q_{3}(x)$ from a family of inverse-gamma densities with degrees of freedom more than $d_{k}$, such as $d_{k}+.5$, and $q_{4}(x)$ similarly with d.f. $d_{k}-.5$. Place the mode of $q_{3}(x), q_{4}(x)$ to occur at $x_{2}^{\mbox{\tiny{max}}}$.
	
To specify $D^{(t)}$, separate into $D^{(t)} \equiv \mbox{min}\left( (1-(A^{t}))\times C^{(t)},1 \right)$, where $C^{(t)}$ is independent of $A^{(t)}$ and $\mathbb{E} \left[ C^{(t)} \right] = \frac{1}{F_{2}}$. 
Simulating $C^{(t)} \sim \frac{q_{3}(X_{2})}{f_{2}(X_{2})}$ where $X_{2}$ comes from $q_{2}(x_{2})$ is sufficient, as long as $(1-A^{(t)})C^{(t)} < 1$ is guaranteed.  If $F_{3<2} < 1$ this will always be the case.   To hold this in all cases, sample $A^{(t)}$ as:
\begin{equation} A^{(t)} \sim \mbox{max}\left( 1- \frac{1}{F_{3<2}}, \mbox{min}\left( \frac{F_{4<2}}{e^{.5}}\frac{f_{2}(X_{4})}{q_{4}(X_{4})}, 1 \right) \right) \mbox{.} \label{DefinitionOfAt} \end{equation}
$X_{4}(t)$ comes from distribution $q_{4}(x)$.  This gives $A^{(t)}$ a strong probability of being $1$, keeping the chain stuck in the on-state, depending on the quality of $q_{4}(X_{4})$.  While  $F_{4<2}$ need not  be a true upper bound, $P(A^{(t)} \neq 1)$  will be improved if $F_{4<2}$ is a close upper bound.  In Section~\ref{Sec:JRSSTheoreticalResults} we give theoretical support to this procedure.

\subsection{Examples}
\begin{figure}[htbp]
	\centering
		\includegraphics[width=0.90\textwidth]{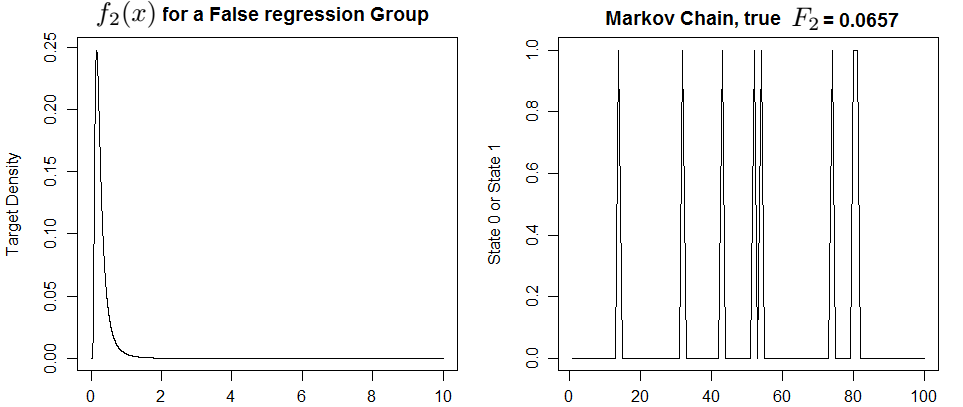}
	\caption{This demonstrates the switching procedure Markov chain when $\int f_{2}(x) dx = .0657$.}
	\label{fig:WorkingMarkovChainBayesSpike}
\end{figure}

  In Figure~\ref{fig:WorkingMarkovChainBayesSpike}, $f_2(x)$ as seen on the left suggests  low evidence for $\tau^2_{k}$, with $\int f_{2}(x)dx = .0657$, making target posterior probability of $ 0.0616 = .0657 / (1.0 + .0657) = F_{2}/(1+F_{2})$.  The chain on the right has 8 on-states out of an initial 100 iterations, and eventually 610 on-states out of 10,000 iterations.

\begin{figure}[htbp]
	\centering
		\includegraphics[width=0.90\textwidth]{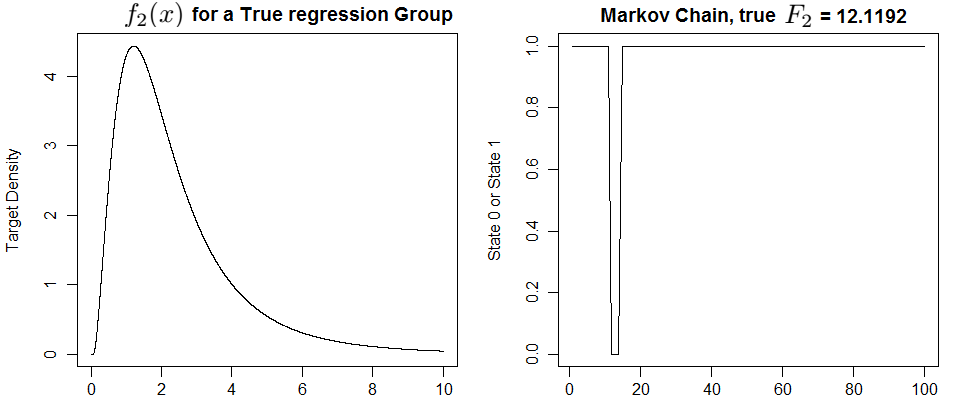}
	\caption{In this chain,  $F_{2} \approx 12.1$ and target is .923.}
	\label{fig:WorkingMarkovChainBayesSpikeFalse12Group}
\end{figure}

When $F_{2} >> 1$ we expect predominantly more on-states.   In Figure~\ref{fig:WorkingMarkovChainBayesSpikeFalse12Group}, $F_{2} = 12.1$ and the sampler is on for 9166 in 10K iterations compared to a target of .923.

\subsection{Theoretical Results}\label{Sec:JRSSTheoreticalResults}
\subsubsection*{\emph{Theorem: $W^{(t)}$ is uniformly ergodic.}}
  $W(t)$ has been constructed such that if $W(t)\sim \mbox{Bernoulli}(F_{2}/(1+F_{2}))$, then $\mathbb{E}^{W(t)}[W(t+1)] = F_{2}/(1+F_{2})$.  We use coupling arguments from~\cite{GeneralStateSpaceSummary}.  Consider one chain $W^{(t)}$ started at $t=0$ arbitrarily on state $0$.  Consider another chain $\tilde{W}^{(t)}$ which is started at convergence: $\tilde{W}^{(0)} \sim \mbox{Bernoulli}(F_{2}/(1+F_{2}))$.  Base future draws of $\tilde{W}^{(t)}, W^{(t)}$ on the same sequence of $\{A^{(t)}, D^{(t)}\}$ draws.  Let $\mathcal{T}$ be the first time $t > 0$ that $\tilde{W}^{(\mathcal{T})} = W^{(\mathcal{T})}$, at which point, couple the chains.  At the times $t < \mathcal{T}$ it will either be that $\{ W^{(t)}, \tilde{W}^{(t)}\} = \{0,1\}$ or $\{1,0\}$, since the chains are uncoupled.  They will couple on the next draw with probability $p^{(t)} \equiv 1 - A^{(t)} - D^{(t)} - 2 A^{(t)} D^{(t)}$, which has expectation 
			
	\begin{equation}  \bar{p}  = 
	1- \mathbb{E} \left[  A^{(t)} + (1-A^{(t)})C^{(t)} + 2A^{(t)}(1-A^{(t)})C^{(t)} \right] \mbox{.}\end{equation}
	
	By nested expectation we can calculate moments for $\mathcal{T}$:
	 \begin{equation}  \mathbb{E} \left[ \mathcal{T} \right] = \frac{1}{\bar{p}} \mbox{, and }
	\mbox{Var} \left[ \mathcal{T} \right] = \frac{1-\bar{p}}{\bar{p}^2} \mbox{ .}\label{CouplingTimeExpectation}\end{equation}
And more importantly, by the coupling theorem we have:
\begin{equation} | 2(P(W^{(t)}=1)-\frac{F_{2}}{1+F_{2}}) | \leq (1-\bar{p})^{t} \mbox{, }\label{CouplingTheorem} \end{equation}
which is exponential convergence, or uniform ergodicity.

The eigenvalues of the transition matrix in Equation~\ref{SwitchingDistributon} are $\lambda_{1} = 1$ and $\lambda_{2}^{(t)}=A^{(t)} - D^{(t)}$ with expected value $\bar{\lambda}_{2} = \mathbb{E} \left[ A^{(t)} \right] - \mathbb{E} \left[ D^{(t)} \right]$.  Using~\cite{Sokal1989} we can find an effective sample size of the chain. For any function$f:W^{(t)} \rightarrow \mathbb{R}$ an autocorrelation function $C_{f}(s)$ defined as:
\begin{equation} C_{f}(s) = \lim_{t \rightarrow \infty} \mathbb{E} \left[ (f(W^{(t)}) - f(F_{2}/(1+F_{2})))(f(W^{(t+s)}) - f(F_{2}/(1+F_{2}))) \right] \mbox{.} \label{Autorcorrelation} \end{equation}
In the case $f(W^{(t)}) = W^{(t)}$,  $C_{W}(s) \approx \exp \{ - s / (-\log_{e} \bar{\lambda}_{2}) \}$.   Define time, $\mathcal{T}_{\mbox{\tiny{int $w$}}}$:
\begin{equation} \mathcal{T}_{\mbox{\tiny{int $w$}}} \equiv \frac{1 + \bar{\lambda}_{2} }{1-\bar{\lambda}_{2}} \label{IntegratedwTime} \end{equation}
from the spectral radius formula (2.36) in~\cite{Sokal1989}.
The number of equivalently independent samples in a converged chain $\tilde{W}^{(t)}$ ran for $n$ steps is then:
\begin{equation} 
  n_{\mbox{\tiny{independent}}} = \frac{n}{2\mathcal{T}_{\mbox{\tiny{int $w$}}} } \mbox{. } 
\label{nindependentequation} \end{equation} 

\section{Implementation}\label{SectionImplementation}
 We implement our method an R-package  ``BayesSpike''.   The  ``Modules'' interface of the ``Rcpp'' R-package~\citep{Rcpp1} allows the R prompt access to C++ class methods.   Algorithm~\ref{alg1} states the sequence of method calls for our Gibbs sampler.
   
   \begin{algorithm}\label{alg1}
   \caption{Group and Fixed Effects Spike Regression $\mathbf{Y} = \mathbf{X} \boldsymbol{\beta} + \boldsymbol{\varepsilon}$}

   \begin{algorithmic}
   \For{$\mbox{Temperatures in }$ $\mathcal{T}\mbox{:}1$}
   
   \For{$\mbox{Chains in }$ $1\mbox{:}\mathcal{C}$}
   
      \textbf{Initalize} random new draws $\boldsymbol{\beta}^{(0)}, \boldsymbol{\tau}^{(0)}, \boldsymbol{\pi}_{\mbox{\tiny{$\mathcal{A}$}}},
        \sigma^2,$
        
     \For{\mbox{$t$ in } $1\mbox{:}\mbox{Length(Chain)}$}
        \If{(\mbox{\textbf{mod}}(t,D) == 0)}   
          \State \verb@EESamplerMerge();@  \emph{\# Jump to $\boldsymbol{\beta}$ at same energy in previous Temp}
        \EndIf
        
        \State \verb@SampleFixedB();@ \emph{\# sample $\hat{B}^{(t)}$ for fixed $\beta_{j}$}
        \State \verb@SampleNewTaus();@ \emph{\# sample $\hat{\tau}^2_{k}$ for groups $\mathbb{J}(k)$}
        \State \verb@RefreshOrderedActive();@ \emph{\#  Allocate new $\mathbf{X}^{T}\mathbf{X}$ columns}
        \State \verb@PrepareForRegression();@ \emph{\# Fill $\mathbf{Q}_{\mbox{\tiny{$\mathcal{A}$}}}^{(t)}$}
        \State \verb@SamplePropBeta();@       \emph{\# Draw 
				$\hat{\boldsymbol{\beta}}^{(t)} $
				        													}
        \State \verb@FillsBetaFromPropBetaAndCompute();@ \emph{\# Recompute $\mbox{XTResid}^{(t+1)}$ }
             
        \If{ROBIT}  \verb@RobitReplace();@ \emph{\# Redraw $\vec{Y}$ for Robit Regression}
        \Else 
         \State \verb@UpdateSigma();@   \emph{\# Update ${\sigma^2}^{(t)}$}
							
         \If{T-NOISE} \verb@UpdateTNoise();@ \emph{\# Redraw weights $\vec{w}$
				  					}
         \EndIf
        \EndIf
        
        \State\verb@RecordHistory();@ \emph{\# Record  and compress 
				  ${\sigma^2}^{(t)}, {\boldsymbol{\tau}^2}^{(t)}\boldsymbol{\beta}^{(t)}$ draws.}
       \EndFor
      \EndFor
     \EndFor 

   \end{algorithmic}
   \end{algorithm}   
\section{Credibility and Assessment}\label{SectionCredibility}
 The purpose of Gibbs samplers is not to produce a loss-minimizing point estimate $\hat{\boldsymbol{\beta}}$ in the least computational time, but to produce actionable risk metrics for our certainty on $\boldsymbol{\beta}$ with respect to draws $\boldsymbol{\beta}^{(t)}$ from the posterior.  These chains produce marginal model inclusion probabilities (MIPs) $\mathbb{E} \left[ B_{j} | Y \right] $ by using Rao-Blackwellized draws, representing either individual coefficient inclusion or group inclusion.  If we wished to assess whether at least one parameter $B_{j_1}, B_{j_2}, \ldots$ in a neighborhood of effects is nonzero, Gibbs samples can be used to calculate
	\begin{equation} \mathcal{P}( B_{j} = 1 \mbox { or } B_{j'} = 1 \mbox{ or } B_{j''} = 1 \ldots |Y  ) \approx \frac{1}{T}\sum_{t} \cup_{\mbox{\footnotesize{$j$ of interest}}} B_{j}^{(t)} \mbox{ . } \label{BjModelInclusion} \end{equation}
	
	We use ``Highest-posterior Density'' (HPD) regions as credibility regions for $\hat{\beta}_{j}$, which should help to produce narrow intervals.  Because $\mathcal{P}(\beta_{j} = 0 | Y )$ has infinite density with respect to our prior,  HPD intervals should, in theory, always include zero.  In practice, $\hat{\beta}_{j}^{(t)}$ draws for a high MIP coefficients are so far away from zero that conventional estimates of HPD intervals, such as in the Coda package~\citep{PlummerCODA}, result in intervals that rarely include zero.  Credibility intervals, while they contain $1-\alpha$ of posterior probability, can lack true frequentist coverage.  Consider $\beta_{j} =  10^{-9}$, unit noise $\sigma^2 = 1$, and a modest sample size $n=100$.  Given other parameter behavior, and our priors for active density $\tau^2_{\mbox{\tiny{$\mathcal{A}$}}} $, MIPs will be near zero for this parameter, causing a $1-\alpha$ HPD interval to be the point interval $(0,0)$.  While close to the truth, this will almost never cover the true $\beta_{j}$.  
	
	For times when it is important that small $\beta_{j}$ estimates are reported with credibility intervals that do have realistic coverage, we propose instead, ``unbounded intervals''.  Consider draws from the posterior:
	\begin{equation} \hat{\beta}_{j}^{\mbox{\tiny{unbounded}}} \sim \mbox{ from density } \mathcal{P}( \beta_{j} | B_{j} =1, \betanj, Y) \label{UnboundedDensity} \end{equation}
	These posterior draws are taken during the usual MCMC algorithm but stored in a separate, non-sparse file-buffer. These are draws where coefficient $j$ is always active but  other coefficients $/j$ have been drawn from a sparse posterior.  Credibility intervals taken from $\hat{\beta}_{j}^{\mbox{\tiny{unbounded}$(t)$}}$ draws should better cover near-zero coefficients.  
	
	\subsection{Default Priors}
	So far, we have detailed an algorithmic implementation for large-$p$ Bayesian posteriors,  but have been agnostic about inputs.  Unlike \emph{lasso}, which requires a single parameter $\lambda$ optimized from cross validation, this Bayesian algorithm suggests a need for priors on parameters $\pia, \sigma^2$ and on the group/or/fixed slab density $\tau^2_{\mbox{\tiny{$\mathcal{A}$}}}$.  When information is significant, $n > p$, or both are modest in size, priors can be uninformative and spread across the space of all models.   But when we must study datasets where $n << p$, we are implicitly assuming  very sparse, relatively large signals.  As defaults, we choose $\pia \sim \mbox{Beta}(1,p)$, $\sigma^2 \sim \mbox{Inv-ChiSquare}(.25 \mbox{Var}(y),n)/n$, and $\tau^2_{\mbox{\tiny{$\mathcal{A}$}}} \sim \sigma^2 / \mbox{Expo}(1)$.  These are based upon a practical assumption: an analysis can only investigate models of size $\mathcal{O}(1)$ without a significant  increase in data.  
	This assumes that the noise level must be smaller than the variation of the data itself, and that non-zero coefficients will be at the same scale as the noise.
	  
\section{Simulation Study}
   We consider several popular penalized-regression estimators and Bayesian MCMC selection estimators, some with settings accounting for prior information.  By prior information, we mean knowledge (potentially approximate or incorrect) of the magnitude of noise level: $\sigma^2$, and the true number of non-zero parameters: $k$, and of the approximate size of non-zero parameters: $\|\beta_{\mbox{\tiny{$\mathcal{A}$}}}\|/\sigma$.  
	
	We use test two priors for Group Bayes. First an ``automatic'', and conservative, choice of $\pia \sim \mbox{Beta}(1,p)$ prior.   If we had knowledge of number of active coefficients, the  ``correct'' prior would be $\pia \sim \mbox{Beta}(k,p-k)$.  We test our automatic prior against  a random, incorrect prior $\pia \sim \mbox{Beta}(k^{\mbox{\tiny{noise}}}, p-k^{\mbox{\tiny{noise}}})$ where $k^{\mbox{\tiny{noise}}} \sim k e^{N(0,3)}$, suggesting that in the real world the number of active coefficients is unknown but will frequently be known to an order of magnitude.  
	For the ``slab'' portion, we use a Gaussian prior $\beta_{j} \sim \mbox{N}(0,\sigma^2 \times \tau^2_{\mbox{\tiny{fixed}}})$ if $\beta_{j} \neq 0$.   Then the prior on $\tau^2_{\mbox{\tiny{fixed}}} \sim 40 / \mbox{Gamma}(40)$, so that posterior update is $\tau^2_{\mbox{\tiny{fixed}}}|\hat{\boldsymbol{\beta}}$ $\sim (\sum_{j}\hat{\beta}_{j}^2/\hat{\sigma}^2 + 40) / \mbox{Gamma}(\|\hat{\boldsymbol{\beta}}\|_{0} + 40)$. For grouped coefficients, we assume an active prior of $\tau^2_{k} \sim 1/\mbox{Gamma}(1)$ independent for each group. 
	
			The Lars (Least Angle Regression) package performs a sweep of \llasso $\mbox{ }$ parameters.  The original $C_p$ minimizer criterion for Lars requires input of $\sigma^2$.  		Elastic Net, is a combination of an $L_{1}$ and an $L_{2}$ penalty term $\lambda_{1}|\beta_{j}| + \lambda_{2} \beta_{j}^2$.  These parameters can be chosen through cross-validation, and we also show Elastic Net performance when we choose models of size $k$ and $k^{\mbox{\tiny{noise}}}$.  As we will see, giving the oracle true model size $k$ to Elastic Net can create an extremely powerful estimator, but inexact knowledge comes at a cost.  The ``SCAD'' penalty~\citep{FanLiSCAD}, which behaves as a \llasso $\mbox{ }$ penalty near zero, but diminishes in penalty further from zero in such a way that the penalized likelihood surface is approximately smooth, is chosen using the ``Minimum-Convexity'' criterion aided by \cite{NCVREG}, as well as the BIC-like criterion , ``WLT'', from ~\citep{Wang01082007} which seems better for small samples. We use minimum-convexity to optimize the similar \cite{zhang2010} Minimax Concave Penalty, ``MCP''.
		
				In these simulations we introduce a sister-method, the 2-Lasso, an EM arg-max estimate which approximates Bayes-B selection using a prior:
		\begin{equation} \beta_{j} \sim \mbox{Laplace}( B_{j} \lambdaa + (1-B_{j}) \lambdad) \mbox{.} \label{2LassoPrior} \end{equation}
		 $\lambdaa$ and $\lambdad$ serve as two competing \llasso $\mbox{ }$ penalties, one which is very restrictive near-zero, and one which is not.  $\lambdaa, \lambdad$ are chosen automatically through equations that maintain an smooth posterior; $B_{j} \sim \mbox{Bernoulli}(\pia)$  a-priori .  In 2-Lasso, the E-step is an update of $\hat{B}_{j}$ which is also a posterior estimate of model inclusion for parameter $j$, and the M-step is Coordinate Descent.
		
		Of Bayesian Gibbs sampler penalties, we will use the \cite{2009arXiv0907MONOMVN} implementation of the Bayes Lasso~\citep{BAYESLASSO} and Horseshoe~\citep{HORSESHOE},  and Spike and Slab~\citep{IshSpikeAndSlab}, run for their default chain-lengths (1000, 250, and $\mbox{n-iter1} = 500, \mbox{n-iter2} = 500)$.  For BayesVarSel~\citep{2016arXiv161108118G}, we choose the highest probability model from 1500 iterations, and choose $\hat{\beta}$ the ML estimate given this model.

	\subsection{Small size simulation}
	We start with an $n>p$ simulation small enough to run against all competing methods. We use 6 randomly-located nonzero (-1, and +1) coefficients, while $\sigma=1.5$, with $n=100$ data-points and $p=25$ total coefficients. $\mbox{Cov}(X_{ij},X_{ik})= \rho^{|k-j|}$ where $\rho=.9$. We perform the experiment 500 times on the UNC Killdevil cluster, which allows reservation of up to 400 cores to individually fit each estimator and simulation. We will report $L_{2}$ error as measured $\sum_{j}(\beta_{j}-\hat{\beta}_{j})^2 / \sum_{j} \beta_{j}^2$.  We report Type 1 and the Type 2 errors as the count of each errors, so if there were $p=25$ coefficients and $6$ are truly non zero, the maximum number of Type 2 errors is 6 and the maximum possible number of Type 1 errors is 19.
	
	In terms of $L_{2}$ error, both ungrouped GB priors come in superior, with cross-validated 2-Lasso coming close.  Here, GB is a choosier estimator than almost all other selectors.  Other selectors choose a large model which always includes truth.  GB makes a small sacrifice in power (missing 3\% of a single coefficient out of six), for the a significant reduction of false positive rate of only 2\%, which is clearly helpful to improved $L_{2}$.   Though GB generated 3 chains of 1000 iterations each, estimation times continue to be competitive.
	   \begin{table}
	\caption{$n=100,p=25,\sigma=1.5$}	
	\centering\begin{centering}
 \begin{tabular}{l|r|r|r|r|} 
    & $L_{2}$ & Type 1 & Type 2 & Time (sec) \\ \hline
  2-Lasso CV & 
     $.181$($.07$) & $.99$($\mbox{1.{\footnotesize36\normalsize}}$) & $.001$($.032$) & $.511$($.032$) \\ 
  \hline Elastic Net CV Min & 
     $.272$($.054$) & $\mbox{4.{\footnotesize79\normalsize}}$($.592$) & $\mbox{0.\footnotesize{0}}$($\mbox{0.\footnotesize{0}}$) & $\mbox{1.{\footnotesize17\normalsize}}$($.085$) \\ 
  Elastic Net oracle k & 
     $.206$($.133$) & $.151$($.364$) & $.177$($.405$) & $.013$($.004$) \\ 
  Elastic Net k-Noise & 
     $.581$($.298$) & $\mbox{5.{\footnotesize03\normalsize}}$($\mbox{6.{\footnotesize57\normalsize}}$) & $\mbox{2.{\footnotesize16\normalsize}}$($\mbox{2.{\footnotesize38\normalsize}}$) & $.012$($.0022$) \\ 
  SCAD minConvex & 
     $.326$($.061$) & $\mbox{8.{\footnotesize52\normalsize}}$($\mbox{2.{\footnotesize37\normalsize}}$) & $.001$($.032$) & $.024$($.0053$) \\ 
  SCAD WLT & 
     $.254$($.083$) & $\mbox{2.{\footnotesize89\normalsize}}$($\mbox{2.{\footnotesize16\normalsize}}$) & $\mbox{0.\footnotesize{0}}$($\mbox{0.\footnotesize{0}}$) & $.101$($.012$) \\ 
  Lars $C_{p}$ & 
     $.311$($.078$) & $\mbox{7.{\footnotesize13\normalsize}}$($\mbox{3.{\footnotesize55\normalsize}}$) & $\mbox{0.\footnotesize{0}}$($\mbox{0.\footnotesize{0}}$) & $.014$($.0013$) \\ 
  MCP minConvex & 
     $.444$($.216$) & $\mbox{13.{\footnotesize0\normalsize}}$($\mbox{5.{\footnotesize29\normalsize}}$) & $.734$($\mbox{1.{\footnotesize97\normalsize}}$) & $.027$($.0034$) \\ 
  \hline \hline GB Prior(1,p) & 
     $.162$($.067$) & $.014$($.118$) & $.022$($.16$) & $.413$($.036$) \\ 
  GB Prior (k-Noise,p) & 
     $.183$($.1$) & $.166$($.489$) & $.071$($.401$) & $.411$($.032$) \\ 
  \hline Ishwaran Spike  & 
     $.241$($.062$) & $\mbox{16.{\footnotesize9\normalsize}}$($\mbox{1.{\footnotesize53\normalsize}}$) & $\mbox{0.\footnotesize{0}}$($\mbox{0.\footnotesize{0}}$) & $\mbox{1.{\footnotesize8\normalsize}}$($.136$) \\ 
  HorseShoe & 
     $.207$($.054$) & $\mbox{16.{\footnotesize5\normalsize}}$($\mbox{1.{\footnotesize45\normalsize}}$) & $\mbox{0.\footnotesize{0}}$($\mbox{0.\footnotesize{0}}$) & $\mbox{2.{\footnotesize07\normalsize}}$($.112$) \\ 
  BayesLasso & 
     $.22$($.056$) & $\mbox{16.{\footnotesize7\normalsize}}$($\mbox{1.{\footnotesize45\normalsize}}$) & $\mbox{0.\footnotesize{0}}$($\mbox{0.\footnotesize{0}}$) & $\mbox{1.{\footnotesize55\normalsize}}$($.118$) \\ 
  BayesVarSel & 
     $.219$($.078$) & $\mbox{1.{\footnotesize09\normalsize}}$($\mbox{1.{\footnotesize13\normalsize}}$) & $\mbox{0.\footnotesize{0}}$($\mbox{0.\footnotesize{0}}$) & $\mbox{3.{\footnotesize02\normalsize}}$($.236$) \\ 
   \hline \end{tabular} \end{centering} \end{table}

	\subsection{Medium size simulation}
	We slightly increase the problem to $p=1000$, while keeping $n=100$, $k=6$, $\sigma=1.5$.  The GB prior is still successful largely due to its low Type 1 error while keeping Type 2 still low.  Bayesian Variance Selection returns a ``A Bayes Factor is infinite'' error, but because BVS seeks to explore the Bayes Factor of the space of all models, we must understand that its specialty is in the cases $p < 100$.   GB completes in 2 seconds.  To be sure, this is only the amount of time to run 3 chains of length $1000$, sufficient to have a stable median $\hat{\boldsymbol{\beta}}$ point estimate.  More iterations should be run to take credibility intervals.  While the arg-min estimates run in less than a second, Bayesian methods expand into the hundreds.  The CV implementation of Elastic net also expands in computation costs.  As we move toward larger $p$, certain models will not be as available to the computation limits of our cluster.	 \begin{table}\caption{$n=100, p=1000$} \begin{centering}
   \begin{tabular}{l|r|r|r|r|} 
    & $L_{2}$ & Type 1 & Type 2 & Time (sec) \\ \hline
  2-Lasso CV & 
     $.26$($.17$) & $.154$($.446$) & $.328$($.717$) & $\mbox{6.{\footnotesize19\normalsize}}$($.627$) \\ 
  \hline Elastic Net CV Min & 
     $.437$($.158$) & $\mbox{5.{\footnotesize38\normalsize}}$($.717$) & $.392$($.74$) & $\mbox{2.{\footnotesize66e3\normalsize}}$($\mbox{1.{\footnotesize44e2\normalsize}}$) \\ 
  Elastic Net oracle k & 
     $.459$($.22$) & $\mbox{1.{\footnotesize02\normalsize}}$($.861$) & $\mbox{1.{\footnotesize13\normalsize}}$($.9$) & $.718$($.077$) \\ 
  Elastic Net k-Noise & 
     $.767$($.181$) & $\mbox{27.{\footnotesize5\normalsize}}$($\mbox{35.{\footnotesize6\normalsize}}$) & $\mbox{2.{\footnotesize59\normalsize}}$($\mbox{2.{\footnotesize46\normalsize}}$) & $.728$($.087$) \\ 
  SCAD minConvex & 
     $.487$($.109$) & $\mbox{10.{\footnotesize7\normalsize}}$($\mbox{2.{\footnotesize87\normalsize}}$) & $.148$($.427$) & $.236$($.045$) \\ 
  SCAD WLT & 
     $.79$($.108$) & $\mbox{44.{\footnotesize9\normalsize}}$($\mbox{4.{\footnotesize47\normalsize}}$) & $.08$($.325$) & $.447$($.057$) \\ 
  Lars $C_{p}$ & 
     $.603$($.108$) & $\mbox{31.{\footnotesize6\normalsize}}$($\mbox{10.{\footnotesize9\normalsize}}$) & $.03$($.171$) & $.932$($.115$) \\ 
  MCP minConvex & 
     $.532$($.082$) & $\mbox{19.{\footnotesize9\normalsize}}$($\mbox{2.{\footnotesize88\normalsize}}$) & $.024$($.166$) & $.208$($.034$) \\ 
  \hline \hline GB Prior(1,p) & 
     $.247$($.185$) & $.026$($.16$) & $.407$($.827$) & $\mbox{2.{\footnotesize78\normalsize}}$($.212$) \\ 
  GB Prior (k-Noise,p) & 
     $.317$($.212$) & $.7$($\mbox{1.{\footnotesize62\normalsize}}$) & $.521$($\mbox{1.{\footnotesize01\normalsize}}$) & $\mbox{2.{\footnotesize95\normalsize}}$($.412$) \\ 
  \hline Ishwaran Spike  & 
     $.649$($.143$) & $\mbox{85.{\footnotesize3\normalsize}}$($\mbox{4.{\footnotesize41\normalsize}}$) & $.23$($.466$) & $\mbox{5.{\footnotesize03\normalsize}}$($.302$) \\ 
  HorseShoe & 
     $.441$($.161$) & $\mbox{1.{\footnotesize84e2\normalsize}}$($\mbox{27.{\footnotesize1\normalsize}}$) & $.042$($.22$) & $\mbox{2.{\footnotesize07e2\normalsize}}$($\mbox{17.{\footnotesize4\normalsize}}$) \\ 
  BayesLasso & 
     $.646$($.184$) & $\mbox{3.{\footnotesize43e2\normalsize}}$($\mbox{69.{\footnotesize4\normalsize}}$) & $.106$($.345$) & $\mbox{2.{\footnotesize05e2\normalsize}}$($\mbox{18.{\footnotesize2\normalsize}}$) \\ 
  BayesVarSel & 
     *(*) & *(*) & *(*) & *(*) \\ 
   \hline \end{tabular} \end{centering} \end{table}

	\subsection{Larger simulation}
	We increase the noise a bit more, and go into the $n=1000, p=100000$ range, still with $\rho=.9$.  We see that the GB still achieves at this range. $*$'s appear when an estimator had no successful completions within an allotment of 24 hours and 20 gigabytes of RAM.    The Elastic Net, when given the true amount ``k=6'', is the best estimator in this simulation.  But exact knowledge of this number is rare, a slight amount of noise in assumption on $k$ and the Elastic Net grows quickly in Type 1 error, and cross-validation did not complete.  Both GB and Ishwaran Spike are much faster algorithms than all arg-min estimators.
	 \begin{table}\caption{$n=1K,p=100K, \sigma=2$} 
   \begin{centering}
    \begin{tabular}{l|r|r|r|r|} 
    & $L_{2}$ & Type 1 & Type 2 & Time (sec) \\ \hline
  2-Lasso CV & 
     $.059$($.021$) & $\mbox{0.\footnotesize{0}}$($\mbox{0.\footnotesize{0}}$) & $\mbox{0.\footnotesize{0}}$($\mbox{0.\footnotesize{0}}$) & $\mbox{9.{\footnotesize7e4\normalsize}}$($\mbox{9.{\footnotesize85e3\normalsize}}$) \\ 
  \hline Elastic Net CV Min & 
     *(*) & *(*) & *(*) & *(*)  \\ 
  Elastic Net oracle k & 
     $.064$($.019$) & $\mbox{0.\footnotesize{0}}$($\mbox{0.\footnotesize{0}}$) & $\mbox{0.\footnotesize{0}}$($\mbox{0.\footnotesize{0}}$) & $\mbox{9.{\footnotesize95e3\normalsize}}$($\mbox{7.{\footnotesize21e2\normalsize}}$) \\ 
  Elastic Net k-Noise & 
     $.74$($.222$) & $\mbox{2.{\footnotesize13e2\normalsize}}$($\mbox{3.{\footnotesize1e2\normalsize}}$) & $\mbox{3.{\footnotesize0\normalsize}}$($\mbox{2.{\footnotesize54\normalsize}}$) & $\mbox{1.{\footnotesize11e4\normalsize}}$($\mbox{1.{\footnotesize01e4\normalsize}}$) \\ 
  SCAD minConvex & 
     $.49$($.015$) & $\mbox{1.{\footnotesize42e2\normalsize}}$($\mbox{7.{\footnotesize46\normalsize}}$) & $\mbox{0.\footnotesize{0}}$($\mbox{0.\footnotesize{0}}$) & $\mbox{1.{\footnotesize22e3\normalsize}}$($\mbox{2.{\footnotesize98e2\normalsize}}$) \\ 
  SCAD WLT & 
     $.767$($.042$) & $\mbox{5.{\footnotesize15e2\normalsize}}$($\mbox{18.{\footnotesize0\normalsize}}$) & $\mbox{0.\footnotesize{0}}$($\mbox{0.\footnotesize{0}}$) & $\mbox{1.{\footnotesize34e3\normalsize}}$($\mbox{2.{\footnotesize74e2\normalsize}}$) \\ 
  Lars $C_{p}$ & 
     $.473$($.059$) & $\mbox{69.{\footnotesize4\normalsize}}$($\mbox{35.{\footnotesize5\normalsize}}$) & $\mbox{0.\footnotesize{0}}$($\mbox{0.\footnotesize{0}}$) & $\mbox{1.{\footnotesize08e4\normalsize}}$($\mbox{1.{\footnotesize19e3\normalsize}}$) \\ 
  MCP minConvex & 
     $.545$($.019$) & $\mbox{2.{\footnotesize23e2\normalsize}}$($\mbox{5.{\footnotesize79\normalsize}}$) & $\mbox{0.\footnotesize{0}}$($\mbox{0.\footnotesize{0}}$) & $\mbox{1.{\footnotesize62e3\normalsize}}$($\mbox{3.{\footnotesize52e2\normalsize}}$) \\ 
  \hline \hline GB Prior(1,p) & 
     $.078$($.04$) & $.284$($.566$) & $\mbox{0.\footnotesize{0}}$($\mbox{0.\footnotesize{0}}$) & $\mbox{2.{\footnotesize73e2\normalsize}}$($\mbox{38.{\footnotesize1\normalsize}}$) \\ 
  GB Prior (k-Noise,p) & 
     $.153$($.139$) & $\mbox{4.{\footnotesize34\normalsize}}$($\mbox{8.{\footnotesize78\normalsize}}$) & $\mbox{0.\footnotesize{0}}$($\mbox{0.\footnotesize{0}}$) & $\mbox{4.{\footnotesize7e3\normalsize}}$($\mbox{6.{\footnotesize06e3\normalsize}}$) \\ 
  \hline Ishwaran Spike  & 
     $.849$($.132$) & $\mbox{4.{\footnotesize61e2\normalsize}}$($\mbox{12.{\footnotesize3\normalsize}}$) & $\mbox{1.{\footnotesize32\normalsize}}$($.965$) & $\mbox{8.{\footnotesize12e2\normalsize}}$($\mbox{1.{\footnotesize09e2\normalsize}}$) \\ 
  HorseShoe & 
     *(*) & *(*) & *(*) & *(*)  \\ 
  BayesLasso & 
     *(*) & *(*) & *(*) & *(*)  \\ 
   \hline \end{tabular} \end{centering} \end{table}

	\subsection{Grouped Coefficients}
	We create active groups of size $5$, values $(+1,+1,0,-1,-1)$.  We use $\rho=.2$, for $p_{g}=20K$ groups or $p=100K$.  Alternate methods include the GrpReg R-package~\citep{GrpReg1,GrpReg2}, the grplasso package~\citep{GRPLASSO}, and the StandGL package~\citep{STANDGL}.  2-Lasso relies upon a simpler EM step, where an indicator $B_{k}$ determines if a group has $\lambdaa$ or $\lambdad$ spread, and perhaps outperforms Group Bayes on Type 1 error, but this cannot provide model inclusion or credibility measures.  StandGL was unable to complete.  Group Lasso was  10x faster while only 2X $L_{2}$ error.  
	\begin{table} \caption{Grouped Model, $n=1K$, $p_{g}=20K$}\begin{centering}
   \begin{tabular}{l|r|r|r|r|} 
    & $L_{2}$ & Type 1 & Type 2 & Time (sec) \\ \hline
  2-Lasso-$\pi_{\mbox{\tiny{$\mathcal{A}$}}}$ Group & 
     $.013$($.0016$) & $\mbox{4.{\footnotesize03\normalsize}}$($\mbox{1.{\footnotesize12\normalsize}}$) & $\mbox{0.\footnotesize{0}}$($\mbox{0.\footnotesize{0}}$) & $\mbox{4.{\footnotesize72e2\normalsize}}$($\mbox{69.{\footnotesize7\normalsize}}$) \\ 
  \hline GrpReg package & 
     $\mbox{1.{\footnotesize48\normalsize}}$($.095$) & $\mbox{11.{\footnotesize0\normalsize}}$($\mbox{1.{\footnotesize64\normalsize}}$) & $\mbox{9.{\footnotesize51\normalsize}}$($\mbox{1.{\footnotesize12\normalsize}}$) & $\mbox{1.{\footnotesize04e2\normalsize}}$($\mbox{11.{\footnotesize6\normalsize}}$) \\ 
  Group Lasso & 
     $.038$($.0028$) & $\mbox{6.{\footnotesize49\normalsize}}$($\mbox{1.{\footnotesize49\normalsize}}$) & $\mbox{0.\footnotesize{0}}$($\mbox{0.\footnotesize{0}}$) & $\mbox{44.{\footnotesize1\normalsize}}$($\mbox{3.{\footnotesize14\normalsize}}$) \\ 
  StandGL & 
     *(*) & *(*) & *(*) & *(*)  \\ 
  \hline \hline GB Prior(1,p) & 
     $.013$($.0015$) & $\mbox{5.{\footnotesize98\normalsize}}$($.128$) & $\mbox{0.\footnotesize{0}}$($\mbox{0.\footnotesize{0}}$) & $\mbox{4.{\footnotesize61e2\normalsize}}$($\mbox{29.{\footnotesize7\normalsize}}$) \\ 
  GB Prior (k-Noise,p) & 
     $.013$($.0014$) & $\mbox{5.{\footnotesize96\normalsize}}$($.189$) & $\mbox{0.\footnotesize{0}}$($\mbox{0.\footnotesize{0}}$) & $\mbox{5.{\footnotesize8e2\normalsize}}$($\mbox{1.{\footnotesize94e2\normalsize}}$) \\ 
   \hline \end{tabular} \end{centering} \end{table}

	\subsection{Logistic Regression}
	Here we test the group prior with $n=400$, $p_{g}=200$, $p=1000$, five of which are active groups of length 5 each assigned $(+1,+1,0,-1,-1)$, again $\rho=.2$, and data generates from a logistic binomial probability.  All algorithms will use settings for logistic functions, the Group Bayes prior will first use a robit df. 9 distribution, but convert to the logistic likelihood through importance sampling.  Though the simulation includes no $\beta_{0}$ intercept, all of the estimators fit this as a free parameter.
	
  Group Bayes tackles this problem but misses out on one coefficient on average.  $2-$Lasso is not robust enough to analyze data in this case.  The StandGL package, which has been discontinued,
	takes  nearly 900 seconds to produce a result,  and the $L_{2}$ error is large, yet Type 1 and Type 2 error might have the best average.  In the weak signals given by many logistic regressors, it is better to have noisy but closer prior information than the $\mbox{Beta}(1,p)$ prior.
	
	 \begin{table}\caption{Logistic, Grouped. $n=400,p=1000$} 
	\begin{centering}
   \begin{tabular}{l|r|r|r|r|} 
  $\mbox{ }$  & $L_{2}$ & Type 1 & Type 2 & Time (sec)\\ \hline
		  2-Lasso CV & 
     *(*) & *(*) & *(*) & *(*)  
		\\ \hline
  \hline GrpReg package & 
     $.955$($.0039$) & $\mbox{99.{\footnotesize4\normalsize}}$($\mbox{8.{\footnotesize46\normalsize}}$) & $\mbox{0.\footnotesize{0}}$($\mbox{0.\footnotesize{0}}$) & $\mbox{1.{\footnotesize62\normalsize}}$($.126$) \\ 
  Group Lasso & 
     *(*) & *(*) & *(*) & *(*)  \\ 
  StandGL & 
     $\mbox{2.{\footnotesize05\normalsize}}$($.328$) & $.206$($.555$) & $.032$($.182$) & $\mbox{9.{\footnotesize21e2\normalsize}}$($\mbox{1.{\footnotesize04e2\normalsize}}$) \\ 
  \hline \hline GB Prior(1,p) & 
     $.536$($.206$) & $\mbox{0.\footnotesize{0}}$($\mbox{0.\footnotesize{0}}$) & $\mbox{1.{\footnotesize31\normalsize}}$($\mbox{1.{\footnotesize37\normalsize}}$) & $\mbox{27.{\footnotesize4\normalsize}}$($\mbox{1.{\footnotesize95\normalsize}}$) \\ 
  GB Prior (k-Noise,p) & 
     $.376$($.181$) & $\mbox{0.\footnotesize{0}}$($\mbox{0.\footnotesize{0}}$) & $.545$($.918$) & $\mbox{29.{\footnotesize3\normalsize}}$($\mbox{3.{\footnotesize25\normalsize}}$) \\ 
   \hline \end{tabular} \end{centering} \end{table}

	\subsection{Credibility Coverage}\label{Credibility Coverage}
	Here in Table~\ref{CredibilityInvestigation}, as an average of $1000$ simulations, in a $\rho=.2$, $n=100$,$p=1000$, $\sigma=1$ setting we set $\beta_{j}$ in multiple sizes, and investigate expected MIP, along with coverage and average width of HPD credibility intervals from the "unbounded intervals" methodology of Section~\ref{SectionCredibility}.  We have 18 non-zero $\beta$  ranging from  $($$-2$,...,$-.01$,$.01$,...$2$$)$, no grouping, and use the $\mbox{Beta}(1,p)$ prior.  Though we target HPDs of credibility $.5$ to $.99$,  coverage is slightly conservative, and their width does not depend on the size of $\beta_{j}$.  A $\beta_{j}$ of size $.5$ is necessary to have beyond 50\% average MIP; those smaller $|\beta_{j}| < .5$ are dwarfed by $\sigma$, as well as the larger coefficients.  Without using the unbounded methodology, 99\% credibility intervals for $\beta_{j}=.01$ can have coverage of only $.16$, and average width of $.07$.  
	 \begin{table}\label{CredibilityInvestigation} \caption{Coverage and $[\mbox{Width}]$ for many $\beta_{j}$ set at relative size against noise $\sigma^2$} \begin{centering}
   \hspace{-25mm} \begin{tabular}{l|r|r|r|r|r|r|r|r|}
  $|\beta|$ &  av. MIP & 0.5 & 0.9 & 0.95 & 0.99 \\ \hline
  0 & .001 & $.541$[$.16$] & $.934$[$.38$] & $.972$[$.46$] & $1.$[$.6$] \\ 
  0.01 & .001 & $.531$[$.16$] & $.934$[$.38$] & $.976$[$.46$] & $1.$[$.6$] \\ 
  0.05 & .001 & $.563$[$.15$] & $.934$[$.38$] & $.972$[$.45$] & $1.$[$.6$] \\ 
  0.1 & .002 & $.519$[$.16$] & $.926$[$.38$] & $.971$[$.46$] & $1.$[$.6$] \\ 
  0.25 & .032 & $.504$[$.16$] & $.886$[$.38$] & $.939$[$.46$] & $.987$[$.6$] \\ 
  0.5 & .516 & $.365$[$.16$] & $.812$[$.39$] & $.897$[$.47$] & $.976$[$.62$] \\ 
  0.75 & .945 & $.468$[$.16$] & $.853$[$.39$] & $.914$[$.47$] & $.973$[$.62$] \\ 
  1 & 1. & $.473$[$.16$] & $.866$[$.39$] & $.9s24$[$.47$] & $.98$[$.61$] \\ 
  1.5 & \mbox{1.{\footnotesize0\normalsize}} & $.475$[$.16$] & $.869$[$.39$] & $.929$[$.46$] & $.981$[$.61$] \\ 
  2 & \mbox{1.{\footnotesize0\normalsize}} & $.453$[$.16$] & $.845$[$.39$] & $.916$[$.46$] & $.981$[$.61$] \\ 
   \hline \end{tabular} \end{centering} \end{table}
	
	\subsection{Equi-energy Behavior}\label{Equi-energy Behavior}
It is a potential flaw to assume a single best model for data.  To test EE tempering, we consider a length $p=500$ model ($\mbox{Cor}(X_{ij}, X_{ij'})=.2^{|j-j'|}$), with $k=6$ non-zero coefficients $[-1,-1,-1,1,1,1]$, and then exactly copy that data such that $X_{i(500+j)} = X_{ij}$ for a $p=1000$ model.  We then initiate at the mode of the first model.  For any given $\hat{\beta}_{j}$ and $\hat{\beta}_{j+500}$, there is an equivalent model where one coefficient is on and the other is off, creating $2^{6}$ possible modes which should equally represent in the posterior. 
There are also modes of $7+$ coefficients where $\beta_{j}+\beta_{j+500}$ cooperate, though activation parameter $\pia$ is roughly $.006$ and every additional coefficient incurs nearly a $-5$ hit to the likelihood. 

In Table~\ref{AMIPSims} we consider 4 samplers, giving $5200$ samples per chain with $200$ burnin, with a single chain per temperature.  
In the first, we never change the temperature, in the next three we follow a temperature sequence, using the preceding temperature to EE-seed a new position every 10 steps.  
Furthermore, for the chains at higher temperatures, we anneal to the next lower temperature every 50 iterations, so that higher temperature estimates relax into a local mode.

\begin{table} \label{AMIPSims} \caption{Mean MIP$[\mbox{s.d.}]$, testing Equi-Energy Tempering}
\begin{centering} 
\hspace{-10mm}
\begin{tabular}{l|r|r|r|r|}	
  Av. MIPs &  $(1)$ & $(1.25,1)$ & $(1.5,1.25,1)$ & $(1.6,1.4,1.2,1)$ \\ \hline
    Max First 6 & .986$[$.059$]$ & .874[.143] & .697[.104] & .655[.076]\\ 
    Med First 6 & .554$[$.297$]$ & .514[.143] & .505[.068] & .493[.062]\\ 
    Min First 6 & .058$[$.152$]$ & .167[.155] & .323[.091] & .341[.08]\\ \hline
    Max $2$nd 6 & .943$[$.151$]$ & .835[.155] & .679[.091] & .661[.08]\\ 
    Med $2$nd 6 & .448$[$.297$]$ & .488[.142] & .496[.069] & .508[.062]\\ 
    Min $2$nd 6 & .015$[$.059$]$ & .128[.144] & .305[.104] & .346[.076]\\ \hline
    Average Zeros & 1.75$[$1.06$]\times 10^{-4}$ & 1.76[1.07$]\times 10^{-4}$ & 1.76[1.06]$\times$ $10^{-4}$ & 1.76[1.07]$\times 10^{-4}$\\ 
    Max $988$ Zeros & $.014[.047]$ & $.014[.049]$ & $.014[.049]$ & $.014[.05]$\\ 
   \hline \end{tabular} 
	
	\end{centering} 
	
	\end{table}

We reason that the chains cannot be fully mixed until MIP for the second $6$ non-zero $\beta_{j}$ parameters becomes equivalent to the first $6$.  While a single sampler at base temperature can escape one mode, this is not sufficient time to explore all possible modes.  We observed that temperature level of $2$ was too high to maintain sparsity in the system.  We see that a temperature level of at least $1.5$ is necessary to generate reliable escape, and that smooth temperature transition is necessary to suppress over-entry of $\beta_{j} = 0$ parameters into the model.
 
\section{Data Applications}
A number of investigations in genetics benefit from modeling group inclusion probabilities. This includes studies in which  which each genetic variable (genes, genomic regions, genomes or genomic combination) under study most naturally corresponds to, for example, an 8-level categorical factor, an increasingly common feature of experiments in model organism genetics \citep{DeKoning2017}. We show how Gibbs group samping informs analysis of an 8-parent diallel mouse cross, as well as a haplotype-based QTL mapping study in rats with 6,333 8-level group variants.

\subsection{Mouse diallel of Collaborative Cross founder strains}
Our method was first designed for the ``diallel'' cross breeding experiment.  In a first-generation cross of $K \geq 3$ inbred strains of a given plant or animal,  each of the two parents contributes a full copy of their identically-paired chromosomes to the offspring, with the exception of sex chromosomes.  Considering a quantitative phenotype $Y$ of the off-spring (such as height or weight), the genetic effects of the parents could be modeled as $a$ additive, $b$ inbred status, $m$ maternal parent of origin, $v$ symmetric cross specific, $w$ anti-symmetric cross specific, and then the sex differential features, denoted by $\Delta^{a}, \Delta^{m}, \Delta^{b}, \Delta^{v}, \Delta^{w}$.  For specimen $i$, with mother $j$ and father $k$, and sex represented by sign $S_{\mbox{\tiny{sex}}} \in \{-1,1\}$, we model:
\begin{equation} \begin{split} Y_{i} & = \mu +\underbrace{a_{j} + a_{k}}_{\mbox{\tiny{additive}}} + \underbrace{m_{j}-m_{k}}_{\mbox{\tiny{maternal}}}  + \underbrace{\mathbf{1}_{j=k} ( \beta_{\mbox{\tiny{inbred}}} + b_{j})}_{\mbox{\tiny{inbred}}} + \underbrace{v_{jk} + S_{j<k}w_{jk}}_{\mbox{\tiny{cross-specific}}}  \\
&+S_{\mbox{\tiny{sex}}}(\Delta_{\mbox{\tiny{sex mean}}} + \underbrace{ \Delta^{a}_{j} + \Delta^{a}_{k}}_{\mbox{\tiny{sex additive}}} +
  \underbrace{\Delta^{m}_{j} - \Delta^{m}_{k}}_{\mbox{\tiny{sex maternal}}} + \underbrace{\mathbf{1}_{j=k}(\Delta_{\mbox{\tiny{inbred}}} + \Delta^{b}_{j})}_{\mbox{\tiny{sex-by-inbred}}} +  \underbrace{ \Delta^{v}_{jk} + S_{j<k} w_{jk}}_{\mbox{\tiny{sex-by-cross}}}) + \varepsilon_{i}\mbox{.}
\end{split} \label{TotalDiallelAdditiveModel}\end{equation} 
If there are $K$ strains, the groups are $K$ additive effects $a_{j} \in a_{1}, \ldots a_{K}$s, as well as $K$ inbreeding $b_{j}$ and $m_{j}$ maternal effects, furthermore $K$ for each of the $\Delta^{a}, \Delta^{m}, \Delta^{b}$ effects, but $K(K-1)/2$ for the $v_{jk}, w_{jk}, \Delta^{w}, \Delta^{v}$ effects.  The fixed effects are $\mu, \beta_{\mbox{\tiny{inbred}}}, \Delta_{\mbox{\tiny{sex mean}}}$.

 Although there are $2K^2$ possible cross breeds, it is rare that the experimenter will be able to breed a balanced sample of size $n \geq 2K^2$ covering all breeds.  This model has $3 + 4K+ 2K^2-10$ coefficients to $\beta_{j}$ (-10 saved using Equation~\ref{constrainedPrior}), so the design matrix $\mathbf{X}$ will never be linearly independent.  
	
\begin{figure}
		\centering
			\includegraphics[width=0.65\textwidth]{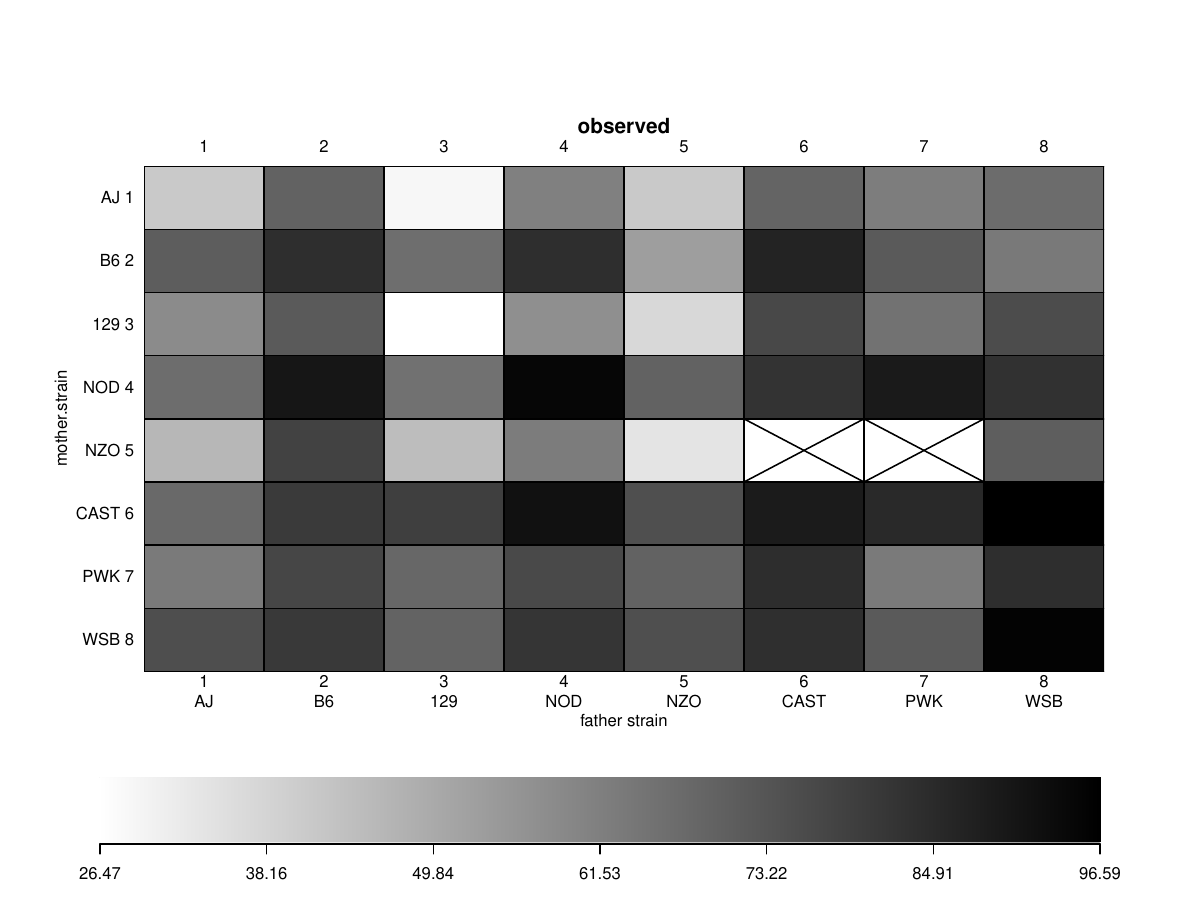}
		\caption{Visualized open field activity in $F_{1}$s of the Collaborative Cross parental strains.}
		\label{fig:OFAPreOnePlotObserved_act_dist_pre1}
	\end{figure}

\begin{figure}[htbp]
		\centering
			\includegraphics[width=1.00\textwidth]{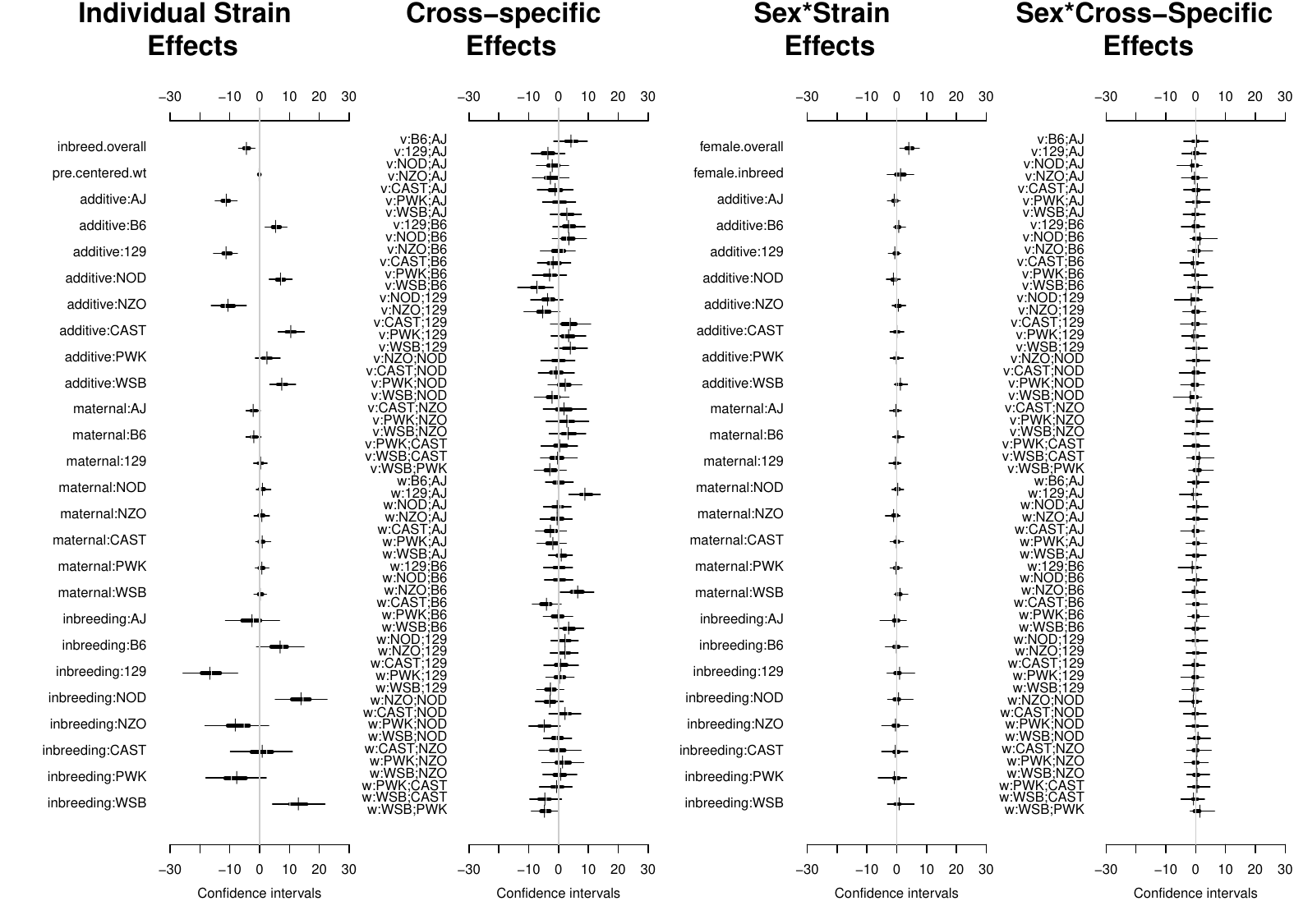}
		\caption{Group Bayes estimated HPD intervals for the effects in a model of Open Field Activity.}
		\label{fig:OFAHPD_act_dist_preCI}
	\end{figure}

	In the ``Collaborative Cross'', $K=8$ inbred mouse strains, (AJ, B6, 129, NOD, NZO, CAST, PWK, WSB) were crossed for multiple generations, so as to mix the genomes and create a wide spectrum of genetic possibilities.  Residual mice from the first generation of the cross reflected draws from an 8-strain diallel, and in \cite{Lenarcic413} and \cite{Crowley321} such diallel mice were tested for phenotypes of genetic variation upon which later generations could be investigated.  Here ``Open Field Activity'' is a measure of total distance traveled for a mouse placed in a $40\mbox{cm}^2$ arena for 30 minutes.  As seen in a visualization of the 8 by 8 breeds in Figure~\ref{fig:OFAPreOnePlotObserved_act_dist_pre1}, this produced noticeable bands marking more active breeds.  As demonstrated with credibility intervals in Figure~\ref{fig:OFAHPD_act_dist_preCI}, the our analysis concluded that one sex parameter, plus additive, inbreed, symmetric and anti-symmetric effects modeled the system, with less evidence for maternal or sex-specific effects.  

\begin{figure}
		\centering
			\includegraphics[width=.8\textwidth]{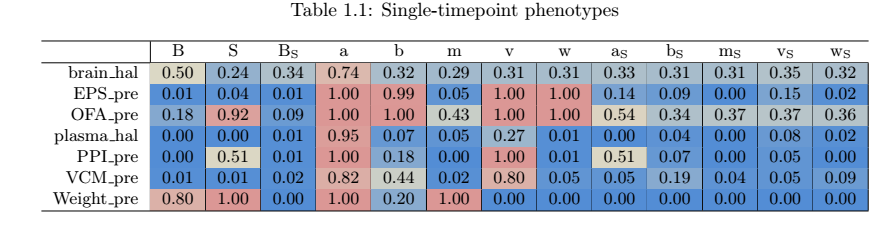}
		\caption{\cite{Crowley321}, Group Bayes estimated Model Inclusion Probabilities for diallel models on ``pre-treatment'' phenotypes.}
		\label{fig:BayesPsychoDiallelMIPs}
	\end{figure}

  Selection by groups informs on the mechanism of heritability for a phenotype. Further analysis is detailed in the~\cite{Crowley321}.  Shown in Figure~\ref{fig:BayesPsychoDiallelMIPs}, MIPs conveniently diagnosed inheritance over many phenotypes.  Measurement variability in diallel phenotypes motivated a model implementing $t$-distributed noise.

\subsection{Heterogenous Stock}

\cite{NatGenRatPhenos} measured 803,485 genotype Single Nucleotide Polymorphisms (SNPs) and 160 phenotypes to identify 230 quantitative trait loci (QTLs) for the Heterogenous Stock Experiment, an intercross population of eight inbred rat progenitors: BN/SsN, MR/N, BUF/N, M520/N, WN/N, ACI/N, WKY/N, and F344/N.  Applying the ``HAPPY'' algorithm of \citet{Mott07112000} to the sequence of SNPs, these binary markers are converted to eight dimensional vectors representing the probability at this position of descent from one of the eight progenitor strains.   We use an additive model of the probabilities for paired chromosomes, and reduce to 6333 markers through reducing to potential loci with at most 95\% correlation with neighbors.  Having a $\tau^2_{k} \sim B_{k}/\mbox{Gamma}(1)$ prior, we use a $\pia \sim \mbox{Beta}(1,(8-1)\times 6333)$ activation prior, include sex as a fixed parameter, and fit a linear mixed model. 

\begin{figure}
		\centering
			\includegraphics[width=.98\textwidth]{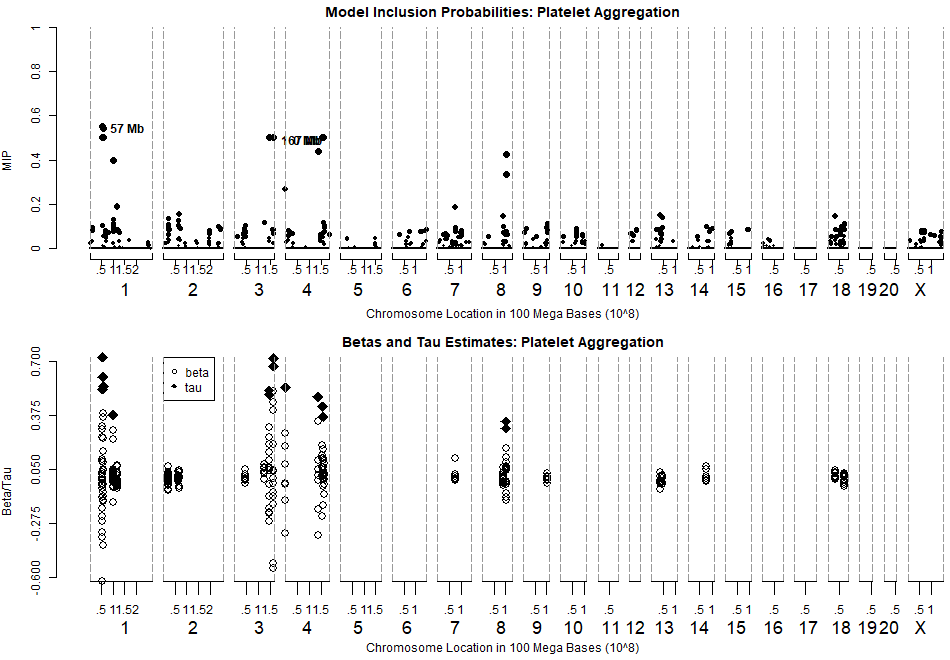}
		\caption{MIP values and $\tau, \beta$ fits for top 15 QTL in Platelet Aggregation}
		\label{fig:BayesHSPlateletAgg}
	\end{figure}

In Figure~\ref{fig:BayesHSPlateletAgg} we repeat the Platelet Aggregation analysis from~\cite{NatGenRatPhenos}, and similarly find a potential QTL near the end of Chromosome 4, though with tentative confidence.  We do find, however, likely additional QTL on Chromosome 1 and 3.  Using the top 15 markers, the posterior mean $\hat{\beta}$ has an $R^2$ of 73\%.

\begin{figure}
		\centering
			\includegraphics[width=.98\textwidth]{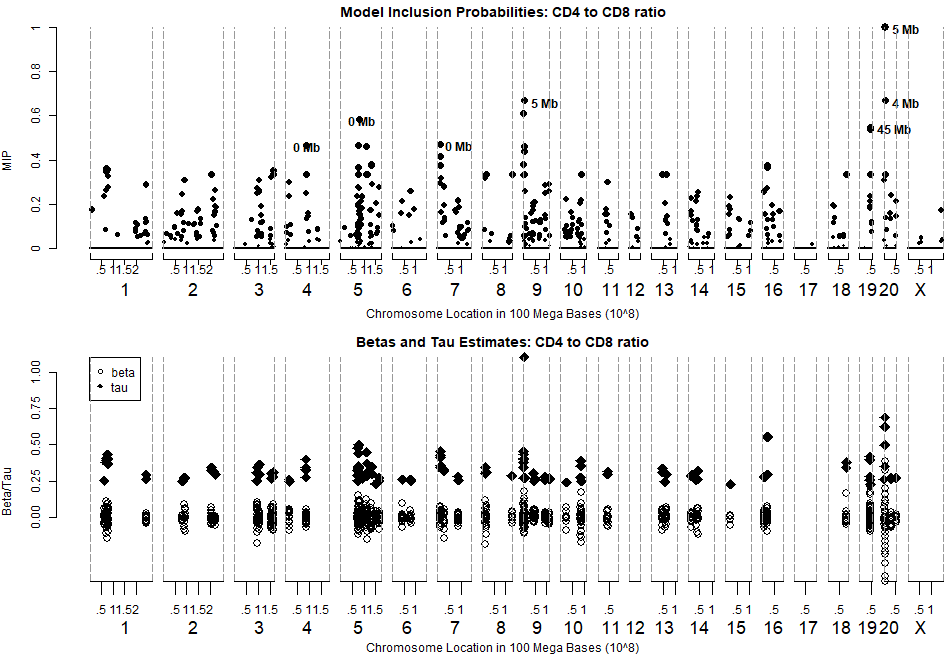}
		\caption{MIP and significant parameter estimates for CD4 to CD8 t-cell ratio}
		\label{fig:BayesHSCD4CD8}
	\end{figure}

For CD4-CD8 ratio, \cite{NatGenRatPhenos} identified QTL at chromosomes 2,9, and 20.  In Figure~\ref{fig:BayesHSCD4CD8} we show results of using a less-restrictive $\pia \sim \mbox{Beta}(1, 6333)$, which permits larger models, having  230 markers with above 10\% model inclusion.  Although markers on chromosome 9 and 20 receive top 10 MIPs, markers on chromosome 2 rank lower than potential candidates on 4,5, and 7.

	Analysis of these phenotypes demonstrates that sparse Bayesian selection is capable of estimation from $p>40,000$ real data, that estimates reflect discoveries from prior methodology, and identify potential routes of new discovery.  Although this Group Bayes procedure can propose new targets from the set of linear mixed models, it cannot so easily grow to add second-order interactions known as epistasis~\citep{EpistasisPhillipsPatrickC2008}, or discover regions of predictable heteroskedacity: termed ``variance QTLs''~\citep{ValdarVQTL}, leaving this one tool for model discovery among many. 

\section{Conclusions}
We have demonstrated an exact method for sparse Gibbs sampling from fixed and random-effects selection distributions, optimized using a unique Markov method to integrate over the collapsed marginal distribution of grouped coordinates.  Using the dynamic reweighting methods of Coordinate Descent, implementing EE tempering, and compressing Gibbs samples, we have ameliorated computational bottlenecks.  As well as showing competitive point-estimate selection against penalized arg-max estimators, this algorithmic approach to sparse Bayes-B/C offers promising confidence measures in MIP and credibility.  Scientific investigations, large and small, benefit from informative, established measures of model confidence.  \bibliography{BayesSpikeRefs}
\end{document}